\documentclass[fleqn,usenatbib]{mnras}

\usepackage{newtxtext,newtxmath}
\usepackage[T1]{fontenc}
\usepackage{graphicx}
\usepackage{amsmath}
\usepackage{amssymb}

\newcommand{\msun}{\ensuremath{{\rm M}_\odot}}

\def\hi{H~{\sc i}}
\newcommand{\nhi}{\ensuremath{N(\mbox{\ion{H}{i}})}}
\def\mgii{Mg~{\sc ii}}

\title[Angular Dependence of the CGM]{Predictions for the Angular Dependence of Gas Mass Flow Rate and Metallicity in the Circumgalactic Medium}

\author[P\'eroux et al.]{C\'eline P\'eroux,$^{1,2}$\thanks{E-mail: celine.peroux@gmail.com} Dylan Nelson$^{3}$, Freeke van de Voort$^{4}$, Annalisa Pillepich$^{5}$, \newauthor
Federico Marinacci$^{6}$, Mark Vogelsberger$^{7}$, Lars Hernquist$^{8}$
\\
$^{1}$European Southern Observatory, Karl-Schwarzschildstrasse 2, D-85748 Garching bei M{\"u}nchen, Germany\\
$^{2}$Aix Marseille Universit\'e, CNRS, LAM (Laboratoire d'Astrophysique de Marseille) UMR 7326, 13388, Marseille, France \\
$^{3}$Max-Planck-Institut f\"{u}r Astrophysik, Karl-Schwarzschild-Str. 1, 85741 Garching, Germany\\
$^{4}$School of Physics and Astronomy, Queen's Buildings, The Parade, Cardiff University, Cardiff, CF24 3AA\\
$^{5}$Max-Planck-Institut f\"{u}r Astronomie, K\"{o}nigstuhl 17, 69117 Heidelberg, Germany\\
$^{6}$Department of Physics and Astronomy, University of Bologna, Via Gobetti 93/2, I-40129 Bologna, Italy\\
$^{7}$Kavli Institute for Astrophysics and Space Research, Department of Physics, MIT, Cambridge, MA, 02139, USA\\
$^{8}$Harvard-Smithsonian Center for Astrophysics, 60 Garden Street, Cambridge, MA, 02138, USA
}

\date{}
\pubyear{2020}

\begin{document}
\maketitle

\begin{abstract}
We use cosmological hydrodynamical simulations to examine the physical properties of the gas in the circumgalactic media (CGM) of star-forming galaxies as a function of angular orientation. We utilise TNG50 of the IllustrisTNG project, as well as the EAGLE simulation to show that observable properties of CGM gas correlate with azimuthal angle, defined as the galiocentric angle with respect to the central galaxy. Both simulations are in remarkable agreement in predicting a strong modulation of flow rate direction with azimuthal angle: inflow is more substantial along the galaxy major axis, while outflow is strongest along the minor axis. The absolute rates are noticeably larger for higher ($\log{(M_\star / \rm{M}_\odot)} \sim 10.5$) stellar mass galaxies, up to an order of magnitude compared to \mbox{$\dot{M} \lesssim 1$ \msun\,yr$^{-1}$\,sr$^{-1}$} for $\log{(M_\star / \rm{M}_\odot)}\sim9.5$ objects. Notwithstanding the different numerical and physical models, both TNG50 and EAGLE predict that the average metallicity of the CGM is higher along the minor versus major axes of galaxies. The angular signal is robust across a wide range of galaxy stellar mass $8.5 < \log{(M_\star / \rm{M}_\odot)} < 10.5$ at $z<1$. This azimuthal dependence is particularly clear at larger impact parameters $b \geq 100$\,kpc. Our results present a global picture whereby, despite the numerous mixing processes, there is a clear angular dependence of the CGM metallicity. We make forecasts for future large survey programs that will be able to compare against these expectations. Indeed, characterising the kinematics, spatial distribution and metal content of CGM gas is  key to a full understanding of the exchange of mass, metals, and energy between galaxies and their surrounding environments.
\end{abstract}



\section{Introduction}

The formation of structure in the Universe is well understood within the standard cosmological framework of $\Lambda$ cold dark matter ($\Lambda$CDM). Small primordial density fluctuations from the cosmic microwave background epoch are amplified through gravitational instability with cosmic time. As fluctuations grow, they create a network of filaments of dark matter and an overdense gaseous medium. This gas ultimately feeds the formation of galaxies, groups and clusters, providing the fuel for star formation. Powerful outflows driven by supernovae and supermassive black holes expel metal-enriched material from galaxies, transforming their surrounding circumgalactic medium (CGM).

These processes of gas motion and conversion are collectively described as the cosmic baryon cycle, which plays out in the CGM. This medium is the interface between the intergalactic medium (IGM) and the interstellar medium (ISM), and contains roughly half of the baryonic mass of the galaxy dark matter halos \citep{werk2014,tumlinson2017}. It not only plays an important role in regulating gas flows into and out of galaxies, but also encodes observational signatures of their baryonic assembly \citep{fg16,chen17}.

Hydrodynamical simulations have shown that the mechanism and morphology of gas accretion onto galaxies are complex \citep{keres05,vandevoort2011a,nelson13}. Gas cooling and inflow can deviate in important ways from spherically symmetric models \citep{silk77,wr78}, most notably by taking the form of filamentary accretion at high redshift \citep{keres05,dekel09,ho19}. At the same time, outflows driven by baryonic feedback processes reverse this flow of baryons into halos.

In simulations, feedback is an integral part of current models, and is needed to explain the most basic properties of the observed galaxy population: the small abundance of low-mass galaxies related to the dark matter halo mass function \citep{vog14a,schaye2015,dubois14,dave16,pillepich18a}; the suppression of star formation or `quenching' in massive galaxies \citep{bower17,donnari19}; the metal enrichment of the ISM, CGM, and IGM \citep{oppenheimer06,muratov17,nelson18b}, and so on. The interaction between inflows and outflows, as well as the ultimate fate of outflowing gas and its ability to escape the halo or recycle back on to the galaxy \citep{fraternali2008,oppenheimer10,angles2017,grand19,nelson19b,mitchell20b} remain open topics of investigation. 

The interaction between a galaxy and its circumgalactic medium is therefore at the centre of our picture into how galaxies form and evolve across cosmic time. However, determining the key physical processes in the CGM remains a difficult task, both because of the complexity of simulating some of the relevant, and small-scale, physics \citep{scannapieco15,mandelker19}, and with the difficulty of obtaining broad and robust observational constraints \citep{nielsen15,rubin18,lehner2019,lan20}. Major open questions include: how do galaxies accrete their gas? How is the CGM structured and enriched with metals? How much mass, energy, and momentum do the galactic winds carry? What is the fraction and physical properties of the material escaping the gravitational potential well? How do inflows and outflows interact with each other?

Due to its diffuse nature (densities n$_{\rm H}<$0.1 cm$^{-3}$), direct detection of the CGM in emission is observationally challenging \citep{bertone2010a,bertone2010b,frank2012,vandevoort2013}, although recent simulations provide ever more robust predictions \citep{augustin2019,wijers2019,corlies20}. In contrast, the circumgalactic medium is commonly detected in absorption against bright background sources \citep{bouche2007a}. This offers a compelling method to study the distribution, chemical properties, and kinematics of the CGM down to relatively low gas densities \citep{peroux2020}. While individual absorption measurements have been limited to a pencil-beam along the line-of-sight, large statistical samples can constrain the mean properties of the CGM around galaxies, showing for example that cool gas is omnipresent around galaxies \citep{adelberger2003, prochaska13,zhu14}. Strong Ly-$\alpha$ absorption has been found around intermediate-redshift $z \sim 2$ galaxies, as well as high covering fractions of metal-enriched gas \citep{adelberger2005, steidel2010, rudie2019}. 

Contemporary to these works, simulations have laid out a refined scheme of the galactic-scale baryon cycle. While gas accretes onto galaxies from the cosmic web filaments, outflowing gas preferentially leaves the galaxy following the path of least resistance, along its minor axis. Where they compete, galactic winds prevent infall of material from regions above and below the disc place \citep{brook2011,mitchell20a, defelippis20}. As a result, inflowing gas is almost co-planar with the major axis of the galaxy \citep{stewart2011, shen2012, vandevoort2012a}. In this canonical view, the accreted gas co-rotates with the central disk in the form of a warped, extended disk, while strong radial/bi-conical outflows are ejected perpendicular to the disk. Observations also show that galactic-scale metal-rich outflows with velocities of several hundred kilometers per second are ubiquitous in massive star-forming galaxies at high redshift \citep[e.g.][]{pettini2001, shapley2003, veilleux2005, weiner2009}. Direct probes of infall are more difficult to obtain, possibly due to the small cross-section or because accretion is swamped by outflows in studies of absorption back-illuminated by the galaxy \citep[the `down-the-barrel' technique; e.g.][]{rubin2014}. When bright background sources are used, projection effects prevent the differentiation of inflowing versus outflowing motion, since redshifted gas could equally be outflowing away from the observer or be infalling gas situated in front of the galaxy \citep{shen2013}. 

Additional information on the orientation of the star-forming disk galaxies is then necessary to help distinguish outflowing from accreting gas. A common approach is to measure the `azimuthal angle' between the absorber and the projected major axis of the galaxy \citep{bouche2012b}. A small (large) azimuthal angle aligns with the major (minor) axis of the galaxy (as illustrated in the left panel of Figure~\ref{fig:tng50_schematic}). In this way, observations have broadly found that absorption is not isotropic -- it depends on the orientation of the galaxy. \cite{bordoloi2011} found a strong azimuthal dependence of \mgii\ absorption, implying the presence of a bipolar outflow aligned along the disk rotation axis. Concomitantly, \cite{bouche2012b} and \cite{kacprzak2012a} found a bimodal distribution of azimuthal angles hosting strong \mgii\ absorption, suggestive of bipolar outflows contributing to the cold gas cross-section \citep{schroetter2019, martin2019}. Such signatures are also found in hotter gas phases including ionised gas traced by OVI absorption \citep{kacprzak2015c}. Interestingly, these trends are not as clearly seen towards active galactic nucleus (AGN)-selected galaxy samples \citep{kacprzak2015a}, implying a stronger connection to stellar feedback. Collectively, these results suggest a picture whereby the cooler component of the CGM at $\sim 10^4$\,K originates from major axis-fed inflows, and/or recycled gas, together with minor axis-driven outflows.

Recently, integral field unit spectrographs (IFUs) have emerged as powerful tools for examining the absorption by gas in the CGM. Early efforts started with the AO-equipped near-infrared spectrograph VLT/SINFONI \citep{bouche2007, peroux2011, peroux2013, schroetter2015, peroux2016}. The potential of this technique for studying the CGM with the wide-field optical spectrograph VLT/MUSE has been further demonstrated \citep{schroetter2016, bouche2016, schroetter2019, zabl2019, muzahid2020, lofthouse2020} as well as with the high spectral resolution optical spectrograph Keck/KCWI \citep{martin19,nielsen2020}. At $z_{\rm abs}<$1, the MUSE-ALMA Halos survey has measured the kinematics of the neutral, molecular and ionised gas of the multi-phase CGM \citep{peroux2017, rahmani2018a, rahmani2018b, klitsch2018, peroux2019, hamanowicz2020}. Large galaxy samples with IFUs increasingly constrain how the physical properties of the CGM vary with angular orientation.

Understanding the galaxy–CGM interactions, i.e., the role of inflows, outflows, star formation, and AGN feedback in governing the gaseous and metal content of galaxies and their environment, is the goal of many recent theoretical efforts. Hydrodynamical simulations of galaxy formation over large cosmological volumes have necessarily limited spatial and mass resolution, which constrains their ability to follow self-consistently the venting of metals by small galaxies and the transport of heavy elements from their production sites into the environment \citep{cen2005, oppenheimer2008, oppenheimer2009, wiersma2009, cen2011}. At the same time, zoom-in simulations with specialized CGM refinement schemes provide enhanced spatial and mass resolution at the cost of galaxy statistics \citep{suresh19, peeples19, hummels19, vandevoort2019}. Over the years, these works have explored the ability of theory to reproduce the main characteristics of the CGM such as ion covering fractions and equivalent width distributions, particularly as a function of impact parameter. We are now able to probe the distribution of metals in the CGM, focusing on how metallicity varies as a function of the galaxy-absorber azimuthal angle.

Our ultimate goal is to map the spatial distribution and characterise the physical properties of the multi-phase gas in the CGM of galaxies. In this paper, we use two recent cosmological hydrodynamical simulations -- EAGLE \citep{schaye2015,crain2015} and TNG50 \citep{nelson19b,pillepich19} -- to investigate how properties of CGM gas vary with azimuthal angle, and develop diagnostics for differentiating inflows and outflows. We provide quantitative predictions for observables, with an eye towards interpreting data that will be available from future surveys. 

The manuscript is organized as follows: Section \ref{sec:methods} presents the simulations and methods used in this study. Section \ref{sec:results} details the global physical properties of the CGM with azimuthal angle, while Section \ref{sec:observables} focuses on observable tracers of these trends. We summarize and conclude in Section \ref{sec:conclusions}. Here, we adopt an H$_0$ = 67.74 km s$^{-1}$ Mpc$^{-1}$, $\Omega_M$ = 0.3089, and $\Omega_{\Lambda}$ = 0.6911 cosmology.


\section{Methods} \label{sec:methods}

In order to obtain robust and more general results, we analyze two distinct cosmological hydrodynamical simulations, based on different numerical methodologies as well as different physical models for baryonic feedback and related processes.

\subsection{The EAGLE Simulation}

The Evolution and Assembly of GaLaxies and their Environments (EAGLE)\footnote{\url{eagle.strw.leidenuniv.nl}} project \citep{crain2015, schaye2015, mcalpine2016} consists of a number of cosmological, hydrodynamical simulations with varying volumes, resolution, and sub-grid physics. The EAGLE simulations were run using a modified version of \textsc{gadget-3} \citep[last described in][]{springel2005}, which is a smoothed particle hydrodynamics (SPH) code. A full description can be found in \cite{schaye2015} and \cite{crain2015} and the EAGLE data releases are described in \cite{mcalpine2016}, but we will briefly summarize the main properties here.

Here, we use the flagship simulation (also referred to as `Ref-L100N1504'), which consists of a cubic volume 100~comoving Mpc on each side with $1504^3$ dark matter particles and as many baryonic particles. The (initial) particle masses for baryons and dark matter are $1.8\times10^6$~M$_\odot$ and $9.7\times10^6$~M$_\odot$, respectively. The gravitational softening length is 0.7~physical kpc at $z<2.8$.

Star formation is modelled following \citet{Schaye2008}. In EAGLE, gas above a metallicity-dependent threshold density (and with temperatures close enough to the prescribed equation of state) is allowed to form stars stochastically, reproducing the observed Kennicutt-Schmidt law \citep{kennicutt1998} by construction. The abundances of 11 elements released by massive and intermediate-mass stars are followed as described in \citet{wiersma09b}. Radiative cooling and heating are computed based on individual elemental abundances, assuming that all the gas is optically thin and in (photo-)ionisation equilibrium with the ionising background radiation from \citet{haardt2001}, as in \citet{wiersma09a}. Feedback from star formation is injected as thermal energy based on the prescription of \citet{dellavecchia2012}. The model for the formation and growth of supermassive black holes and their associated feedback is described fully in \citet{rosas2015} and \citet{schaye2015}. 

The EAGLE simulations were tuned on the $z=0$ stellar mass function and on galaxy sizes (the stellar component), but were not adjusted to reproduce any specific properties of the gas either within or outside of galaxies. They do, however, match a number of observational results fairly well, such as the H\,\textsc{i} column density distribution and covering fraction \citep{rahmati2015} and the distribution and optical depth of metal-line absorption systems \citep{rahmati16, turner16}. Discrepancies with observations of the intergalactic or circumgalactic medium have also been identified, including a lower optical depth of high-ionisation metals and a lower cosmic density of these ions at z$\sim$0 and z$\sim$4 \citep{schaye2015, rahmati16, turner16}.

\subsection{The IllustrisTNG Simulations and TNG50}

The IllustrisTNG\footnote{\url{www.tng-project.org}} project \citep{naiman18, pillepich18b, nelson18a, marinacci18, springel18} is a series of three large cosmological volumes, simulated with gravo-magnetohydrodynamics (MHD) and incorporating a comprehensive model for galaxy formation physics. All data from TNG has been or will be publicly released \citep{nelson19a}.

\begin{figure*}
	\includegraphics[width=7.0in]{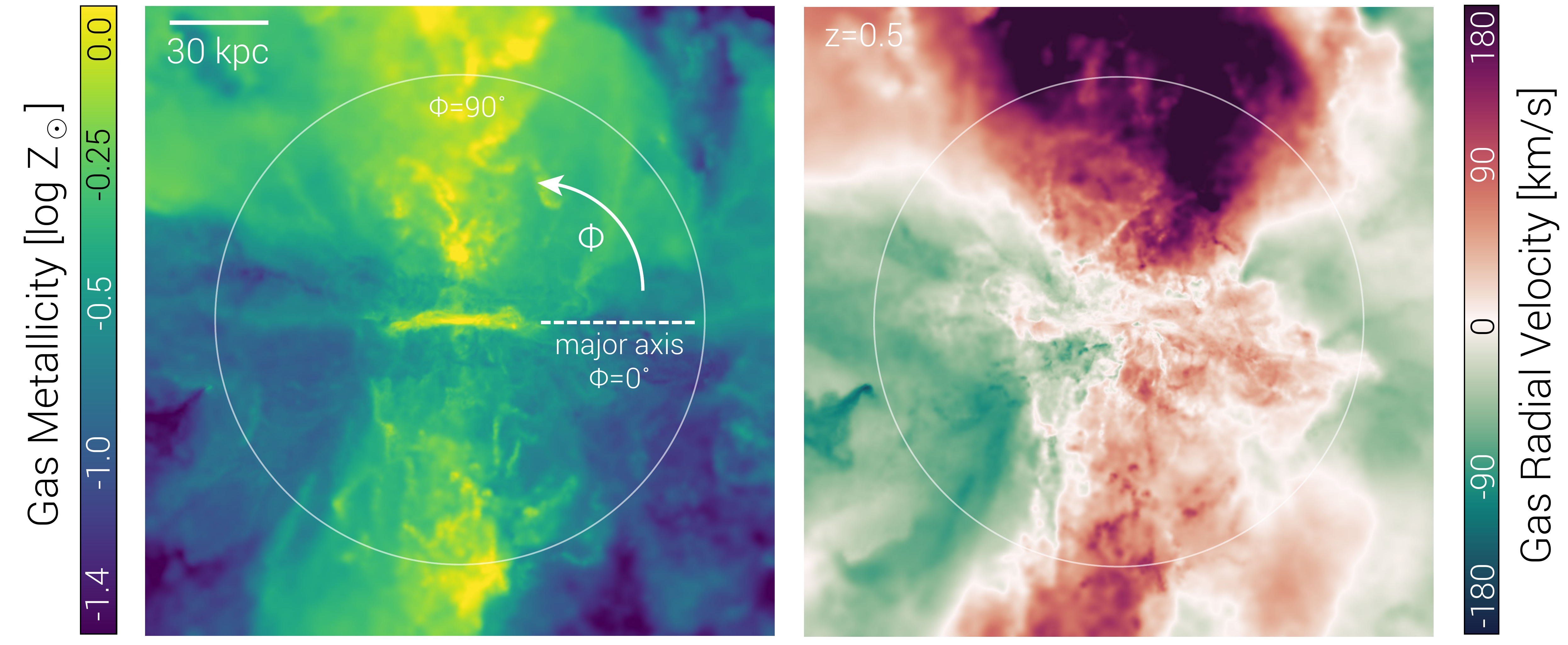}
    \caption{Example of a TNG50 galaxy at $z=0.5$ with a stellar mass of M$_{\rm star}=10^{10.3}$\msun\ and total halo mass of M$_{\rm halo}=10^{11.8}$\msun, rotated edge-on. The azimuthal angle with respect to the galaxy disk is indicated schematically, while the circle shows, $r_{\rm vir}/2$ $(r_{\rm vir}$=150 pkpc). \textbf{Left Panel:} Color shows gas metallicity, demonstrating the overall trend that we find: gas with the highest metallicities preferentially aligns with the minor axis. \textbf{Right Panel:} Color indicates radial velocity (positive denoting outflow; negative denoting inflow), showing that the highest CGM gas metallicities are associated with high-velocity outflowing gas ejected from the interstellar medium, while lower metallicities are associated with accreting/inflowing gas.}
    \label{fig:tng50_schematic}
\end{figure*}

The physical model has been comprehensively described \citep{weinberger17,pillepich18a} and we will only summarize its main properties here. TNG uses the \textsc{Arepo} code \citep{spr10} which self-consistently evolves a cosmological mixture of dark matter, gas, stars, and black holes as prescribed by self-gravity coupled to ideal, continuum MHD \citep{pakmor11,pakmor13}. The physical processes included in the simulations are: gas radiation processes, including primordial and metal-line cooling, plus heating from a meta-galactic background radiation field \citep{faucher09}; star formation within the cold component of a two-phase interstellar medium model \citep{springel03}; the evolution of stellar populations and subsequent chemical enrichment, including Supernovae Ia, II, and AGB stars (independently tracking the ten elements H, He, C, N, O, Ne, Mg, Si, Fe, and Eu); galactic-scale outflows generated by supernova feedback \citep[see][]{pillepich18a}; the formation and mergers of supermassive balck holes (SMBHs) and their accretion of neighbouring gas \citep{springel05a, dimatteo05}; SMBH feedback that operates in a dual mode with a thermal `quasar' mode for high accretion rates and a kinetic `wind' mode for low accretion rates \citep[see][]{weinberger17}. TNG300 is in statistical agreement with the frequency of high-ionisation absorbers at all columns, while TNG100 has slightly too common high column absorption. IllustrisTNG also recovers the covering fractions of all galaxy types, although it does not find a signal in the differential covering fractions of isolated versus non-isolated haloes as claimed by observations \citep{nelson18b}. The TNG50 volume produces sufficiently high covering fractions of extended, cold gas as seen in data. Quantitative comparisons of the predicted low-ionisation column densities with observations reveal no significant tensions \citep{nelson20}.

Here we exclusively use the TNG50 volume \citep{nelson19b,pillepich19}, the highest resolution incarnation. TNG50 includes 2$\times$2160$^3$ resolution elements in a $\sim$\,50 Mpc (comoving) box, giving a baryon mass resolution of $8.5 \times 10^4$\msun, a collisionless softening of 0.3 kpc at $z$\,=\,0, a minimum gas softening of 74 comoving parsecs, and adaptive resolution for the hydrodynamics down to a few tens of parsecs \citep[see Figure 1 of both][]{pillepich19, nelson20}.

\subsection{Galaxy Identification, Sample, and Analysis} \label{sec:analysis}

In both simulations we identify galaxies and the dark matter halos they reside in with the \textsc{Subfind} \citep{spr01} and friends-of-friends \citep[FoF;][]{davis85} algorithms, respectively. Within each FoF halo, subhalos are identified as gravitationally bound collections of resolution elements. In this work we consider only central galaxies, whose positions are taken as the location of the particle/cell with the minimum gravitational potential energy.

We define the stellar mass of a galaxy as the sum of the gravitationally bound stellar particles within an aperture of 30 physical kpc. For the galaxies considered herein, this differs negligibly from other commonly employed definitions, such as the mass within twice the stellar half mass radius. At a given redshift and for a given stellar mass range, we include all central galaxies in the simulations, making no selection on properties beyond $M_\star$\footnote{In the EAGLE sample we also require a minimum of 100 star-forming gas particles in order to better resolved disks.}.

We restrict our analysis to simulated galaxies which are reasonably well resolved. In TNG50 we consider galaxies down to a minimum stellar mass of $10^{8.5}$\msun, which corresponds at $z=0.5$ to $\approx5000$ star particles. In EAGLE we consider galaxies down to a minimum of $M_\star = 10^{9.3}$\msun, having $\approx2500$ star particles.

We derive CGM metallicities with mass-weighted projections of gas particles/cells using the standard adaptively-sized cubic spline kernel onto a grid with pixel scale 0.25 pkpc through a line of sight depth of 500 pkpc. Typically, a galaxy of stellar mass M$_{\rm star}=10^{10}$\msun\ has $r_{\rm vir}$=150 pkpc (defined with respect to the critical density, $r_{\rm 200c}$). In all cases, we rotate galaxies edge-on by diagonalizing the moment of inertia tensor of the star-forming gas; the azimuthal angle for every pixel is defined with respect to this rotated frame. Unless noted otherwise, all gas phases (density and temperature) are included.


\section{Results} \label{sec:results}

Figure~\ref{fig:tng50_schematic} presents a visual representation of an example TNG50 galaxy at $z=0.5$ with a stellar mass of M$_{\rm star}=10^{10.3}$\msun\ and total halo mass of M$_{\rm halo}=10^{11.8}$\msun, rotated edge-on. This is a typical galaxy at this mass scale in TNG50, showing strong gaseous interactions with its surrounding medium in the form of accreting and outflowing gas. This typical object illustrates how the inflowing gas is dominantly co-planar with the major axis of the galaxy, while strong radial/bi-conical outflows are ejected perpendicular to the disk. The left panel of Figure~\ref{fig:tng50_schematic} also presents the definition of the azimuthal angle between the background line of sight and the projected galaxy's major axis on the sky.


\subsection{Gas Mass Flow Rate}

\begin{figure*}
	\includegraphics[width=1.0\columnwidth]{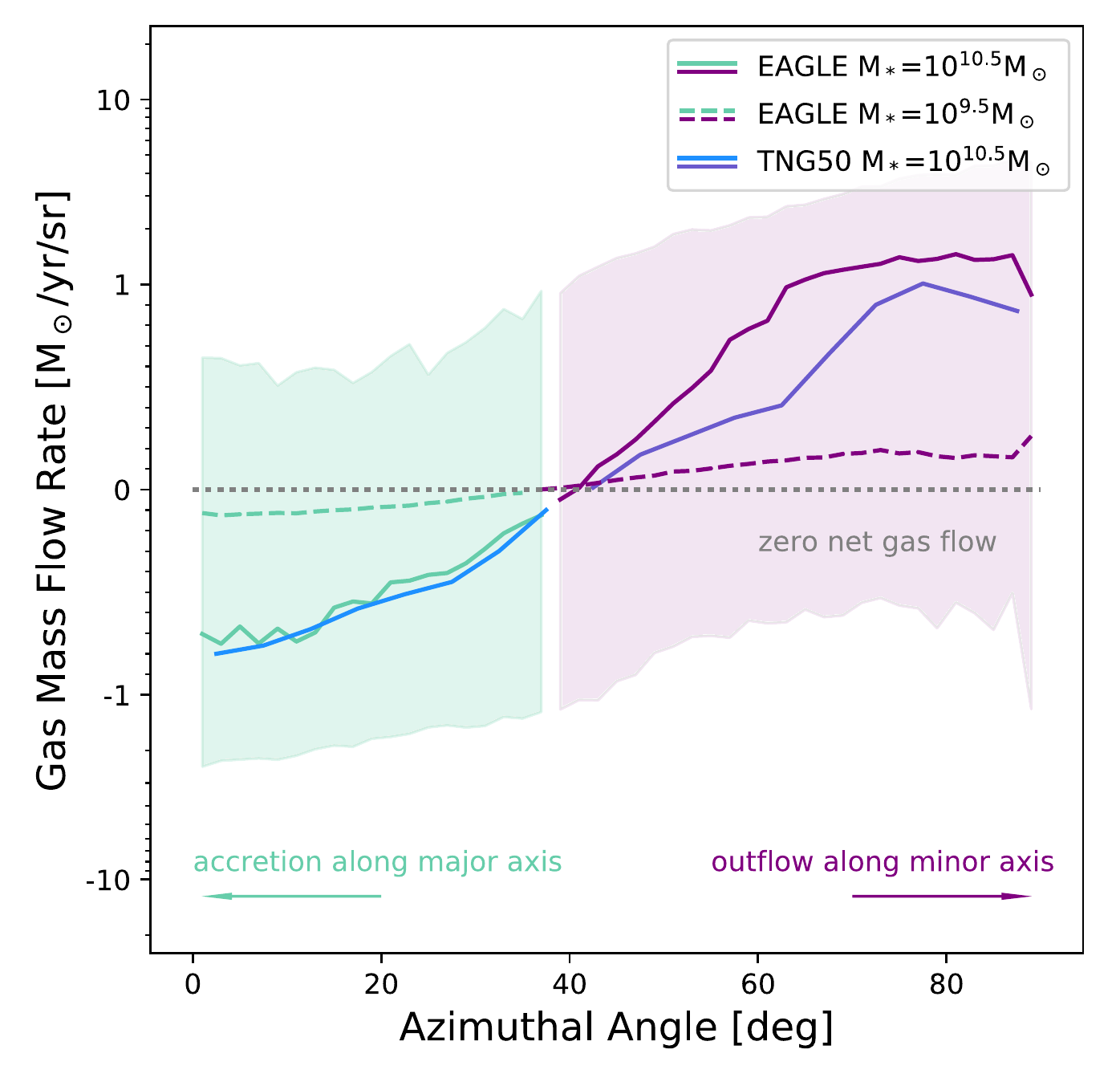}
 	\includegraphics[width=1.0\columnwidth]{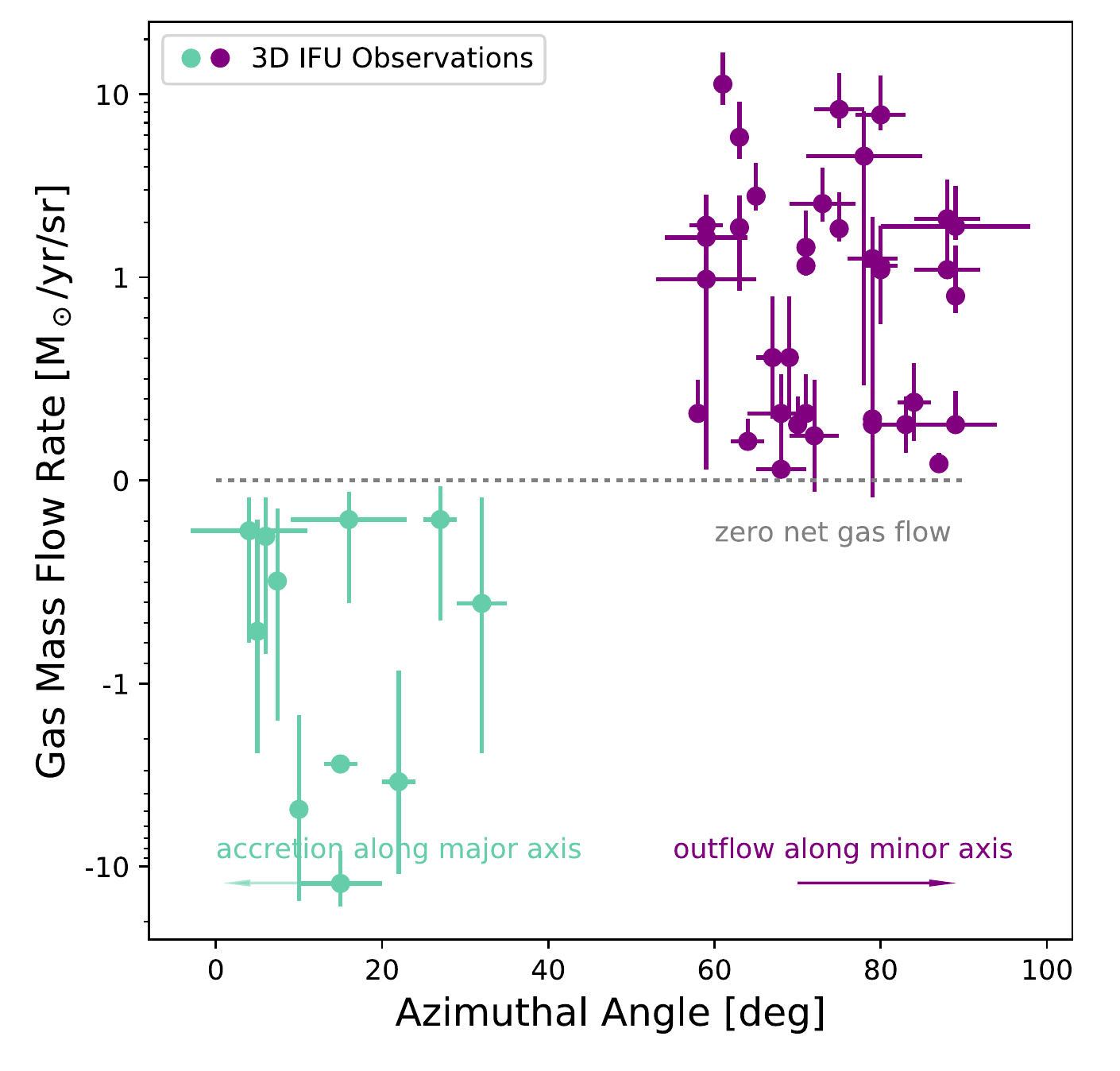}
 	\caption{{\bf Left Panel:} TNG50 and EAGLE predictions of the gas mass flow rate $\dot{M}$ (in units of M$_{\odot}$/yr/sr) as a function of azimuthal angle, defined as the \textit{net} rate of outflow versus inflow at each angle. We stack galaxies similar to quasar absorber hosts detected in current surveys ($z=0.5$, stellar mass of  \mbox{$M_\star = 10^{9.5 \pm 0.2}$ M$_{\odot}$} and \mbox{$M_\star = 10^{10.5 \pm 0.2}$ M$_{\odot}$}, at an impact parameter of 100$\pm$10 kpc). The band shows 16--84$^{\rm th}$ percentiles (i.e. 1$\sigma$, for EAGLE $M_\star = 10^{10.5 \pm 0.2}$ M$_\odot$ only). The horizontal dotted line represents a gas mass flow rate of zero. Note the y-scale is linear at $<\pm$ 1M$_{\odot}$/yr/sr and logarithmic at $>\pm$ 1M$_{\odot}$/yr/sr. TNG50 and EAGLE simulations show a remarkable agreement predicting a clear trend of the inflowing gas along the major axis (negative gas mass flow rates) and outflowing gas along the minor axis (positive gas mass flow rates). {\bf Right Panel:} The data points indicate observational measurements of mass inflow rates by \protect\cite{bouche2013, bouche2016, rahmani2018a, zabl2019}, and mass outflow rates by \protect\cite{bouche2012b, schroetter2015, kacprzak2014, schroetter2019}. We stress that observational works distinguish inflows from outflows based on the azimuthal angle such that the divide along the abscissa of the figure is an assumption of these measurements. The phases probed are also different, as the observations are based on \mgii\ absorption which traces cool, $\sim 10^4$\,K gas, while the simulation results include all gas. Barring a direct comparison at present, our results indicate the expectation that both the direction and rate of gas flow are strong functions of azimuthal angle.}
    \label{fig:mass_flow_rate}
\end{figure*}

The most important property of gas flows in the CGM is the rate of mass outflow or inflow, $\dot{M}$. The mass accretion rate provides a direct description of the gas supply for star formation in the system, constraining the efficiency of star formation. Likewise, the mass outflow rate provides a direct measurement of the ejective efficiency of combined, galactic-scale feedback processes.

By relating the mass outflow rate to the star formation rate (SFR) of the galaxy, the ultimate goal is to determine the mass loading factor of these objects, $\dot{M}_{\rm out}/SFR$. The mass-loading factor characterises the amount of material involved in a galactic outflow, and is an important input to many theoretical models for galaxy formation -- hydrodynamical, analytical, and semi-analytical.

We therefore use the two simulations and measure the instantaneous average gas mass flow rates around galaxies at $z=0.5$. We compute the radial velocity (with respect to the central galaxy) multiplied by the mass of all particles within the shell (of the cone) and divided by the width of the shell. We then normalise the quantity to 1 steradian (sr) rather than the surface area of the cone (which would vary depending on the azimuthal angle). We compute the local gas mass flow rate at an impact parameter $b$ as the average within a spherical shell of $90-110$\,kpc. This shell is taken in 3D rather than 2D projection. We consider two stellar mass bins of \mbox{$M_\star = 10^{9.5 \pm 0.2}$ M$_{\odot}$} and \mbox{$M_\star = 10^{10.5 \pm 0.2}$ M$_{\odot}$}. The redshift of z=0.5 is chosen to match those of on-going MUSE surveys of quasar absorbers \citep{schroetter2019, hamanowicz2020, muzahid2020} while the stellar mass corresponds to the mean stellar mass of high column density quasar absorbers \citep{augustin2018}.

The left panel of Figure~\ref{fig:mass_flow_rate} shows gas mass flow rates as a function of azimuthal angle wrapped every 90 degrees. Negative gas mass flow rate values denote inflows and positive values denote outflows. Our measurement from the simulations represents the \textit{net} rate at each angle, such that an equal flux of inflow and outflow produces zero net, as occurs at an azimuthal angle of $\sim 40$ deg. We find a clear trend whereby inflowing gas dominates along the major axis, while outflowing gas dominates along the minor axis. The correlation is equally present at impact parameter of 50$\pm$10 kpc (not shown for clarity). This signature of gas outflow rate modulated by azimuthal angle is similar to the findings in \citet{nelson19b} for TNG50 at higher redshift.  We remark that \citet{mitchell20a} compare outflow rates between TNG and EAGLE and report notable similarities and/or differences depending on stellar mass and distance from the galaxy. Specifically, while the outflow rates we measure at $M_\star = 10^{10.5 \pm 0.2}$ M$_\odot$ are qualitatively similar between the two models, significant differences exist at other masses (and radii) due fundamentally to the distinct physical models invoked for galactic outflows. Therefore, while the trend reported in Figure~\ref{fig:mass_flow_rate} depends in large parts on the feedback mechanisms implemented in the two different simulations, the results displayed show a remarkable agreement. These findings lend quantitative support to the canonical scenario where the physical properties of the CGM are a strong function of the azimuthal angle. 

On average, absolute gas mass flow rates are small, tending to be constrained to \mbox{$\dot{M} \lesssim 1$ \msun\,yr$^{-1}$\,sr$^{-1}$}. The inflows and outflows have comparable values within a stellar mass bin. The absolute rates are noticeably larger for higher stellar mass galaxies, up to an order of magnitude for the outflow rate. For these galaxies with $M_\star \sim 10^{10.5}$\msun\, there is a strong modulation of net flow rate with azimuthal angle: inflow is strongest at angle$ = 0^\circ$, while outflow is strongest at angle$ = 90^\circ$. The halo-to-halo scatter, shown by the shaded regions, is large and some haloes show much stronger inflow and outflow rates. In a minority of systems, inflows (outflows) can even be found along the minor (major) axis. 

Gas mass flow rates are challenging to measure observationally, partially because the quantity itself is a time derivative, while empirical estimates are fundamentally instantaneous measurements. More critically, the direction of the flow is largely unconstrained, and assumptions must be made about distance, gas column, ionisation state, velocity profile, geometry/opening angle, filling factor, and so on \citep{bouche2013}. Nonetheless, there are a number of gas mass flow rates estimated, predominantly using \mgii\ absorbers, based on a measured impact parameter coupled to a gas flow velocity inferred from toy models which aim to reproduce the spectral characteristics of the absorption profiles.

Relatively small samples at $z=1$ and $z=2$ from the SINFONI instrument \citep{bouche2012b, bouche2013, schroetter2015} have been enlarged using MUSE \citep{bouche2016, rahmani2018a, schroetter2019, zabl2019}. Taken together, these observations tentatively indicate that the mass ejection rate ($\dot{M}_{\rm out}$) is similar to the SFR \citep{schroetter2016, schroetter2019}. At the same time, outflow speeds ($\sim$100 km/s) are generally smaller than the estimated local escape velocity, which could suggest that the outflows do not escape the galaxy halo and are recycled back into the galaxy \citep[although see][]{steidel2010}.

\begin{figure}
	\includegraphics[width=1.0\columnwidth]{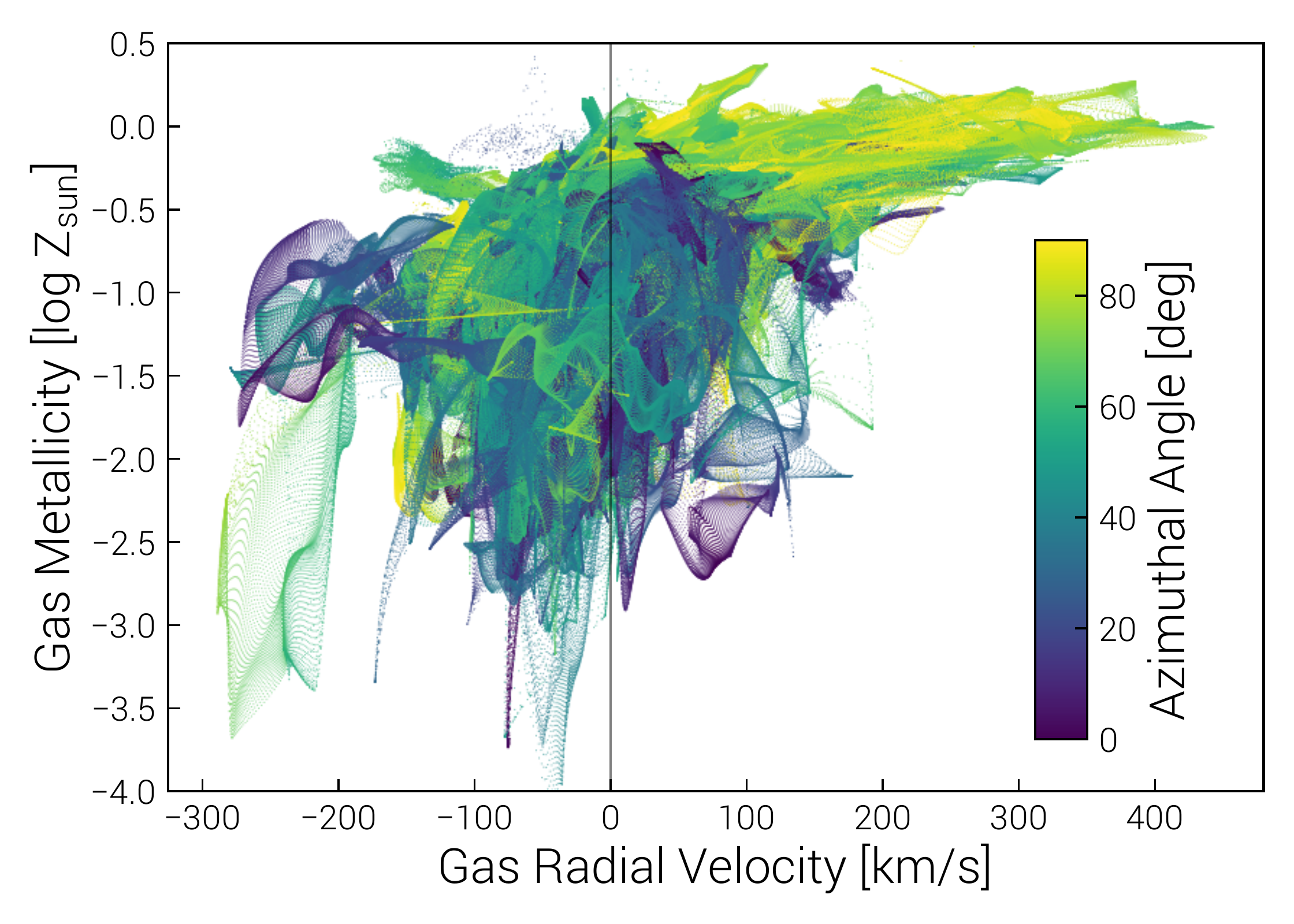}
    \caption{Gas metallicity as a function of radial velocity, with the vertical line demarcating inflow ($<$ 0) versus outflow ($>$ 0). We stack together $\sim 70$ galaxies from TNG50 at $z=0.5$ with stellar mass  $M_\star \simeq 10^{10 \pm 0.1}$\msun, and include all sightlines with an impact parameter 100$\pm$10 kpc, showing that outflows have preferentially higher metallicity than inflows, and occur more aligned with the minor axis of galaxies (larger azimuthal angles, as indicated by the color).}
    \label{fig:Z_vs_velocity}
\end{figure}

In the right panel of Figure~\ref{fig:mass_flow_rate} we also include these observational measurements, converted to units of \msun\,yr$^{-1}$\,sr$^{-1}$ for bi-conical outflows with mean opening angles of 30 degrees. The mass inflow rates as a function of azimuthal angle are taken from the works by \cite{bouche2013, bouche2016, rahmani2018a, zabl2019}, while the mass outflow rates are taken from \cite{bouche2012b, schroetter2015, kacprzak2014, schroetter2019}. We stress that these observational works distinguish inflows from outflows based on the azimuthal angle itself, such that the divide along the abscissa of the right panel of Figure~\ref{fig:mass_flow_rate} is an assumption, rather than a result, of these empirical measurements. The statistics of current observations are relatively limited, and error estimates on individual flow rates are large, such that significantly larger samples and better constraints on the modeling assumptions would be useful to further improve our understanding of the dependence of $\dot{M}$ on azimuthal angle.

Comparing the simulation prediction with the data we note a reasonable agreement in the broadest sense. We caution, however, that the values shown from the simulations represent the average behavior in an azimuthal bin: observational measurements intersect a much smaller subset of CGM gas, restricted to an observable phase, which could result in substantially different values and larger scatter. Observations tend to have higher gas mass flow rates, although this may be a selection effect: current detection limits mean that gas mass flow rates $\dot{M}<$0.05 \msun\,yr$^{-1}$\,sr$^{-1}$ are challenging to detect. The phases probed are also different, as the observations are based on \mgii\ absorption which traces cool, $\sim 10^4$\,K gas, while the  simulation results include all gas, which is generally dominated by hotter, ionised gas. We have explicitly checked that calculating gas mass flow rates including only gas with temperature $T < 10^{4.5}$\,K in the simulation results in low inflow rates ($\dot{M}\sim$0.01 \msun\,yr$^{-1}$\,sr$^{-1}$) for all azimuthal angles, with scatter significantly larger than the azimuthal angle signal.  We find that while there is outflowing cool gas in some of these haloes, it does not dominate the median net gas mass flow rate. Interestingly, \cite{defelippis20} report that there is always inflowing cold gas along the disk plane in TNG100 galaxies, but only high-angular momentum objects show the presence of net outflowing cold gas along the minor axis. Barring a direct comparison at present, our results indicate the expectation that both the direction and rate of gas flow are strong functions of azimuthal angle.


\subsection{Gas Metallicity}

A key diagnostic of the CGM is the metallicity of the gas. Many processes, including supernovae and black hole-driven winds, as well as metals stripped or ejected from satellites galaxies, all leave chemical imprints in the CGM. Their relative contributions, however, involve complex baryonic processes that are challenging to model, and which are poorly constrained by present data. Broadly speaking, gas accreting inwards is expected to have characteristically lower metallicity than gas being expelled outwards.

For example, simulations from \cite{shen2013} show that gas inflows are enriched to metallicities of only a tenth solar by previous episodes of star formation in the main host and in nearby dwarfs. \cite{hafen2020} further speculate that low-ionisation absorption systems are likely to probe accreting gas. At the same time, cosmological models of galaxy formation show that strong outflows are necessary to regulate galaxy growth \citep{angles2014, hopkins2014, somerville2015} and explain the metal enrichment of the IGM \citep{aguirre2001, oppenheimer06}. Indeed, metals are produced in stars within galaxies over billions of years, which enrich the outflowing gas \citep{vandevoort2012b, shen2013}. Therefore, metal-rich gas is likely tracing galactic winds, recycled outflows and gas tidally stripped from galaxies. 

Here, we make a quantitative assessement of this scenario. We compute the gas metallicity as a function of radial velocity, where negative velocities trace inflows while positive velocities trace outflow. We stack together $\sim 70$ galaxies from TNG50 at $z=0.5$ with stellar mass $M_\star \simeq 10^{10 \pm 0.1}$\msun, and include all sightlines with an impact parameter 100$\pm$10 kpc. The results are presented in Figure~\ref{fig:Z_vs_velocity} which shows that high metallicity regions are found preferentially in outflowing gas  and occur more aligned with the minor axis of galaxies, i.e. at larger azimuthal angles. This finding demonstrates the connection between gas metallicity and flow direction. Namely, that outflows have preferentially higher metallicity than inflows, making gas-phase metallicity a valuable indicator for the baryon cycle \textit{phase} gas is participating in. With this scenario in mind, and given the relation established between gas flow direction and azimuthal angle, one naturally expects a trend of increasing CGM metallicity with increasing azimuthal angle, two readily observable quantities.

\begin{figure*}
	\includegraphics[width=6.4in]{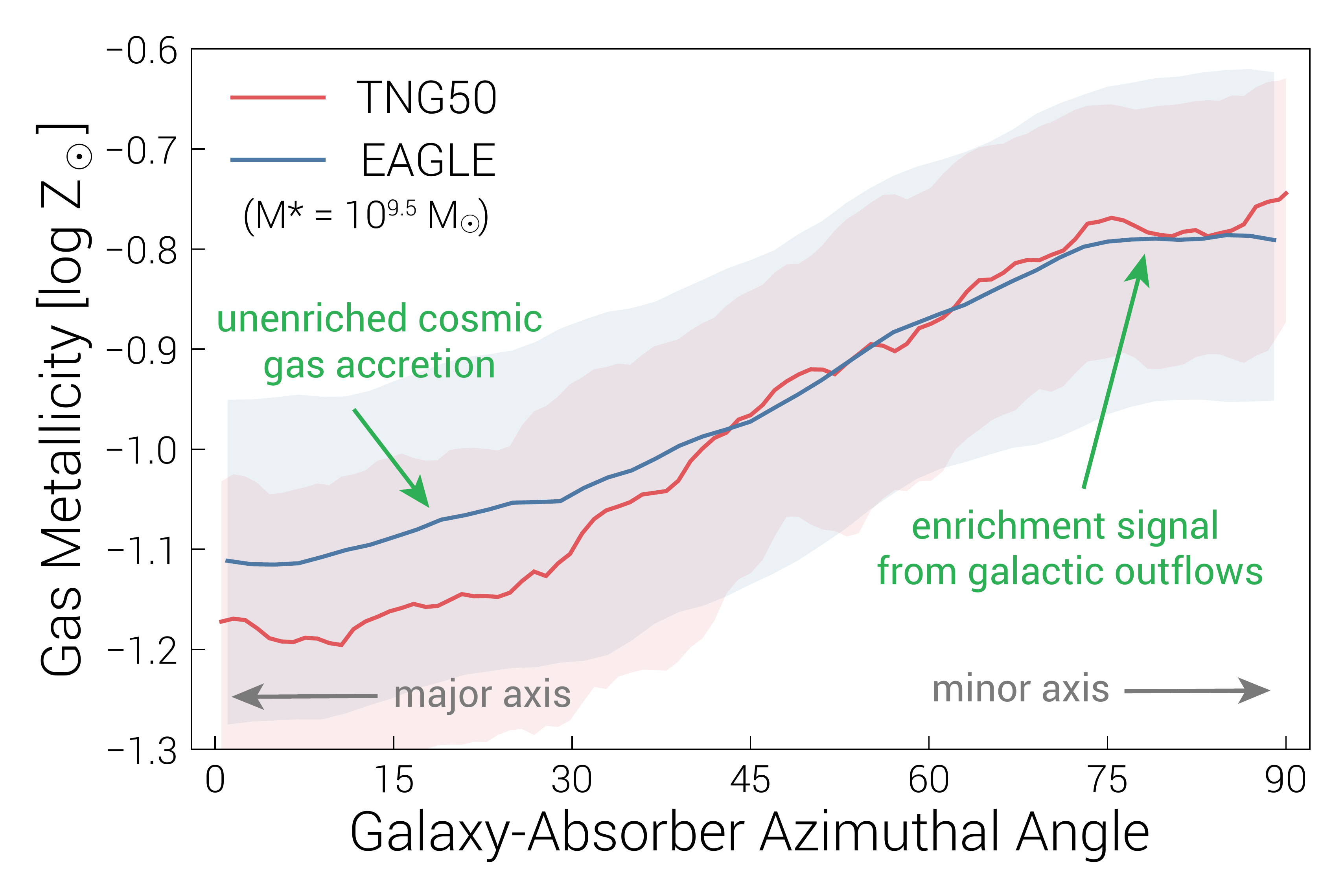}
    \caption{Predictions from two galaxy formation simulations, TNG50 and EAGLE, for the relation between CGM metallicity and azimuthal angle. Here, we stack $\sim$100 ($\sim$500) galaxies in TNG50 (EAGLE) in a narrow stellar mass bin at $10^{9.5 \pm 0.1}$\,\msun, at $z=0.5$ and at impact parameters of $100 \pm 10$ physical kpc. The bands show 38-62th percentiles (i.e. $\sim \sigma/2$) for clarity. There is a remarkable agreement between EAGLE and TNG50 predictions of a correlation between metallicity, Z$_\odot$, and azimuthal angle. Our quantitative predictions are in line with expectations from the canonical view of metal-poor gas accreting co-planar with the major axis of the galaxy while enriched gas is expelled from galaxy along the minor axis.}
    \label{fig:Z_vs_angle}
\end{figure*}

\subsubsection{Fiducial Galaxy Sample and Analysis}
\label{sec:fiducial}

Here, we use both EAGLE and TNG50 (see Section~\ref{sec:methods}) to explore the spatial variation of metallicity in the circumgalactic medium. We consider mass-weighted metallicity, with all gas phases contributing equally. We therefore avoid uncertainties related to estimates of particular elemental abundances or ionisation states, which are affected by models for stellar yields and details of ionisation post-processing including assumptions on collisional versus photo-ionisation from cosmic and local radiation fields.

We define a set of fiducial analysis parameters, which are: a narrow stellar mass bin at {$\pm M_\star = 10^{9.5 \pm 0.1}$\,\msun, redshift $z=0.5$, impact parameter $b = 100 \pm 10$ pkpc}, and $N_{\rm HI,min} = 0$, i.e. no cut-off in \nhi\ column density so that the metallicity summed over all phases of the gas is taken into account. As in Figure~\ref{fig:mass_flow_rate}, the stellar mass range is representative of quasar absorber hosts in current observational samples \citep{peroux2019,hamanowicz2020}. See Section~\ref{sec:analysis} for more details on the analysis. 

Figure~\ref{fig:Z_vs_angle} presents the dependence of CGM metallicity with azimuthal angle, under our fiducial set of analysis parameters. We find a significant trend of increasing metallicity from $Z_{\rm gas}/\rm{Z}_\odot \sim 10^{-1.2} - 10^{-0.8}$ as sightlines shift from the major to minor axis. There is a remarkable agreement between EAGLE and TNG50 predictions. We stress that both the trend and normalisation of the metallicity curves are direct outputs of the simulations, and result from the convolution of the galaxy assembly, stellar enrichment, and baryonic feedback processes in each model. 

These simulations show that CGM regions are subject to non-negligible metal recycling and hence mixing processes over time \citep{nelson20}. Sightlines through the CGM are therefore likely to intersect gas of multiple origins. Indeed gas that accretes onto a galaxy can later be ejected back into the CGM by galactic winds, while gas which was ejected from halos can later re-accrete onto the CGM. Therefore, metal-rich outflows recycle through the CGM and mix with the accreting metal-poor gas \citep{oppenheimer2010, rubin2012, angles2017, zheng2017, hafen2019, hafen2020}. Despite the complexity of the physical processes at play, Figure \ref{fig:Z_vs_angle} shows a consistent picture where CGM metallicity is a function of azimuthal angle. 

\subsubsection{Signal Dependence on Galaxy and Analysis Parameters}
\label{sec:parameters}

\begin{figure*}
	\includegraphics[width=3.4in]{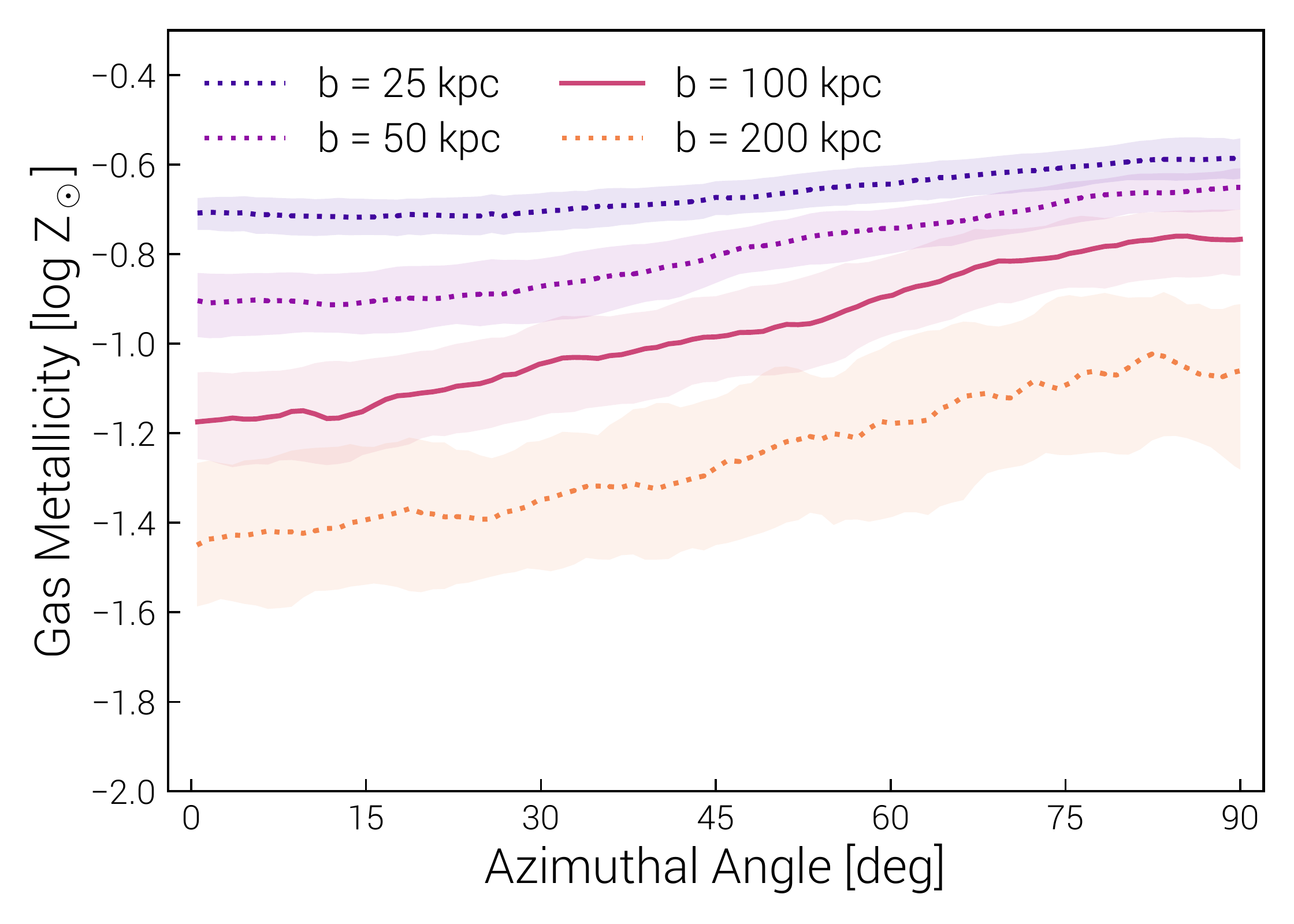}
	\includegraphics[width=3.4in]{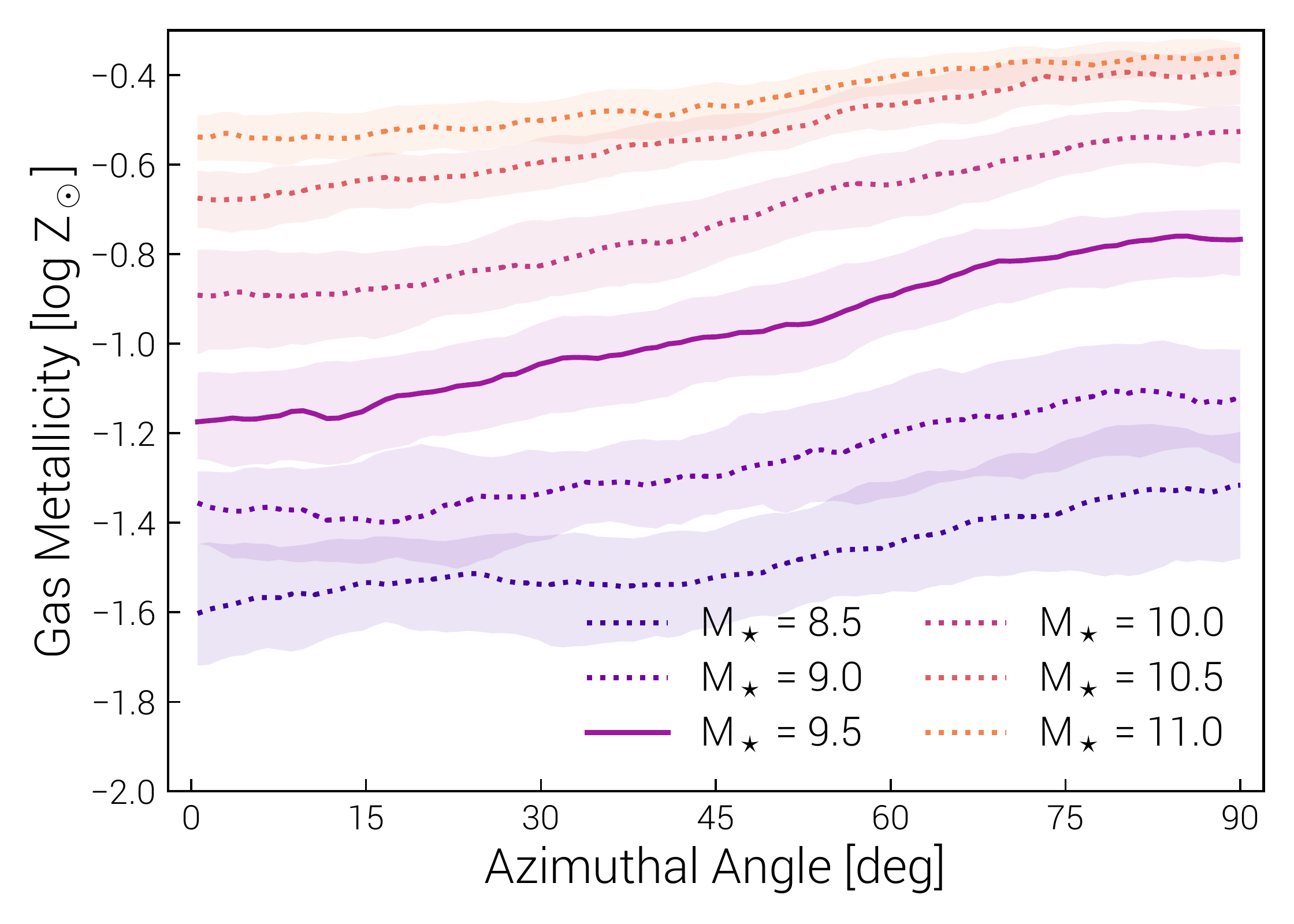}
	\includegraphics[width=3.4in]{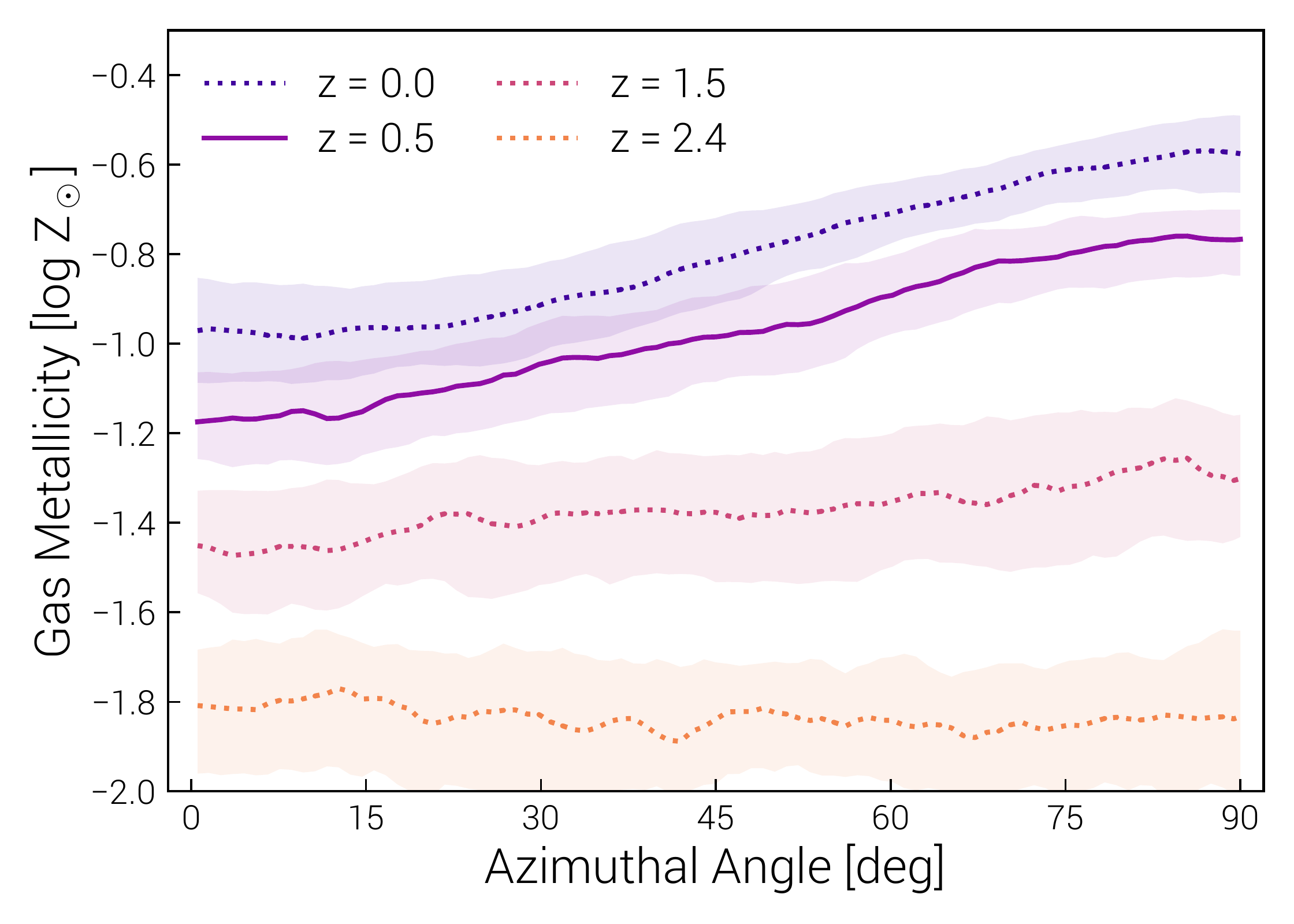}
	\includegraphics[width=3.4in]{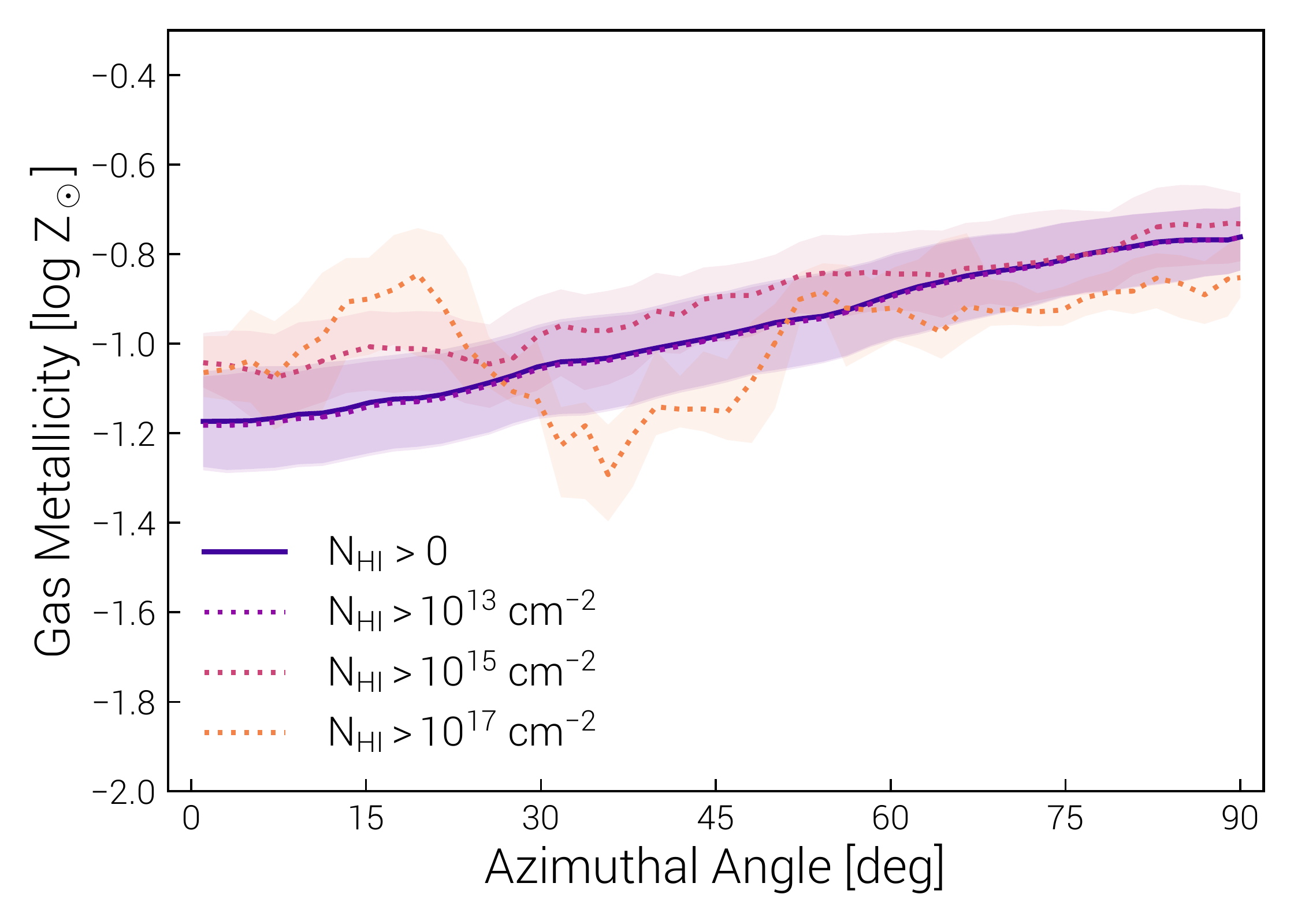}
    \caption{Dependence of the CGM metallicity vs azimuthal angle trend in TNG50 with four parameters: impact parameter (upper left), stellar mass (upper right), redshift (lower left), and $N_{\rm HI}$ threshold (lower right). When not varied, other parameters remain at their fiducial values: $M_\star = 10^{9.5 \pm 0.1}$\,\msun, $b = 100 \pm 10$ pkpc, $z=0.5$, and $N_{\rm HI,min} = 0$ indicated by solid lines. The bands show 38-62th percentiles (i.e. $\sim \sigma/3$) for clarity. The overall metallicity normalization varies strongly with impact parameter, stellar mass and redshift. Importantly, however, the trends of increasing metallicity with azimuthal angle are robust against different parameter choices, including the case when only sightlines with high neutral gas column densities are included.}
    \label{fig:Z_vs_angle_dependencies}
\end{figure*}

We now use the statistics available in the cosmological volume of TNG50 to explore the effects of our choice of fiducial model on predictions of CGM metallicity with azimuthal angle. In particular, Figure~\ref{fig:Z_vs_angle_dependencies} explores trends with impact parameter (upper left), stellar mass (upper right), redshift (lower left), and \nhi\ neutral gas column density threshold (lower right). Each panel varies one parameter at a time, holding the others fixed at our fiducial choices; the solid lines show this same, fiducial result.

The upper left panel of Figure~\ref{fig:Z_vs_angle_dependencies} considers four values of impact parameter: $b=25$, $50$, $100$ and $200$\,pkpc. The most striking result is the change in the overall metallicity normalization, whereby most metal-rich gas is located closer to the galaxy. From just outside the disk to $b \sim 100$\,pkpc, the correlation with azimuthal angle increases with impact parameter. We speculate that this is because fountains do not yet promote metal mixing over the full volume (i.e. range of azimuthal angles) at these distances, as occurs closer to the galaxy. At the largest impact parameter, $b=200$\,pkpc, the differential signal is largely the same as at 100\,pkpc, which is beneficial as observational statistics are typically maximal at these distances.

The upper right panel of Figure~\ref{fig:Z_vs_angle_dependencies} shows the trends for five bins of stellar mass, from $M_\star = 10^{8.5 \pm 0.1}$\,\msun\ to $10^{11 \pm 0.1}$\,\msun. Naturally, we find that the normalisation of CGM gas-phase metallicity is a strong function of the galaxy stellar mass, reflecting the mass metallicity relation \citep{nelson18a,torrey19}. To check whether the effect is partially due to the chosen impact parameter at fixed physical scale, we perform identical calculations at a fixed halo radius instead (r$_{\rm vir}/2$). The results show that while the metallicity normalisation are higher, the offsets with stellar mass
   remain. Importantly, the change of normalization with $M_\star$ generally exceeds the change with azimuthal angle. As a result, observational samples in a sufficiently small, and measured, stellar mass range are needed to uncover the predicted signal. This is particularly challenging given that galaxy stellar mass measurement is not immediately available from an absorption detection, but must be obtained with additional follow-up in the continuum. 

From the theoretical point of view, this broad galaxy stellar mass range encompasses different physical mechanisms for driving galaxy outflows. In the TNG model in particular, these include supernovae-driven winds at lower masses \citep{pillepich18a} and supermassive black hole (SMBH) driven winds at higher masses \citep{weinberger18}. Both feedback mechanisms have been shown to produce anisotropic outflows around TNG50 galaxies \citep{nelson19b}. In the most massive bin shown, $M_\star = 10^{11 \pm 0.1}$\,\msun, SMBH-driven outflows dominate, but the qualitative signal of CGM metallicity versus azimuthal angle remains present, implying that black hole feedback in massive disks can produce a similar signature as supernovae feedback in galaxies one hundred times less massive.

The lower left panel of Figure~\ref{fig:Z_vs_angle_dependencies} presents the redshift evolution of the metallicity trend with azimuthal angle, at $z=2.5$, $1.5$, $0.5$ and $z=0$. We recover the general evolution of increasing metallicity with cosmic time \citep{decia2018, poudel2018}, as stellar populations age, more stars are formed and enrich the Universe towards redshift zero. The signal of metallicity with azimuthal angle increases towards the present day: at high-redshift the metallicity gradient is shallower than at $z=0$. These results indicate that observational samples should target lower redshift objects when possible, and that we expect no angular signal whatsoever, for our fiducial impact parameter and stellar mass bin, at $z \gtrsim 2$ where the cosmic time is too short to enrich the CGM.

Finally, the lower right panel of Figure~\ref{fig:Z_vs_angle_dependencies} evaluates the impact of including only sightlines above a threshold (minimum) neutral gas column density $N_{\rm HI}$. This is because observational determination of the metallicity requires measurement of the amount of hydrogen present, in addition to the amount of a metal tracer. Observations focus on the cold (i.e. neutral hydrogen) phase at $\sim 10^4$\,K where ionisation corrections are minimal, implying that sightlines with metallicity measurements could probe biased regions of the CGM. We find that the trend of metallicity with azimuthal angle remains qualitatively present for most $N_{\rm HI}$ thresholds. 

This is not obvious a priori, as many sightlines with lower neutral hydrogen column densities pass through primarily more diffuse and hotter regions of the CGM, including unobserved phases at $\gtrsim 10^6$\,K. The trends become less statistically significant with higher $N_{\rm HI}$ cuts, and larger samples are required to recover the signal. At the highest column thresholds (lighter-colour lines), the trend with azimuthal angle becomes non-monotonic as a peak emerges near a zero azimuthal angle. As this aligns with the disk plane, we speculate that this could be due to enrichment of inflows from fountain mixing within the CGM, and/or the preference of accreting satellite galaxies and their associated denser, enriched gas to be found aligned with the angular momentum axis of the central.

We note that to reduce the noise in this panel due to the relative sparsity of high $N_{\rm HI}$ column sightlines we increased the size of the impact parameter bin from $100 \pm 10$ to $100 \pm 30$ pkpc. We also carry out two further checks: (i) computing the signal including only gas cells with a neutral gas metallicity of $\gtrsim 10^{-3}$, and (ii) computing the signal where the contribution of each gas cell is weighted by neutral \hi\ mass instead of total gas mass. The result of (i) is similar to the results we show varying $N_{\rm HI,min}$, and the result of (ii) is qualitatively consistent with all the trends we present, only tending to slightly increase the overall metallicity normalisation independent of azimuthal angle.

Overall, Figure~\ref{fig:Z_vs_angle_dependencies} demonstrates that the result of increasing metallicity with azimuthal angle is robust with choices of different parameters. Furthermore, the normalization varies strongly with impact parameter, stellar mass and redshift. While observational samples with well-determined impact parameter and redshift are readily available, determination of the galaxy stellar mass is less common. Indeed, measures of the stellar mass of the galaxy associated with the gas probed in absorption requires detection of the galaxy continuum over a wide wavelength range. It is clear that a sample spanning too large a range in $M_\star$ will fail to clearly distinguish the underlying trend, and that homogeneous samples are needed. The angular signal is, however, commonly found in many accessible regimes, making it a robust and generic feature of lower-redshift galaxies. 


\section{Discussion: Implications for Observables} \label{sec:observables}

\begin{figure}
	\includegraphics[width=1.0\columnwidth]{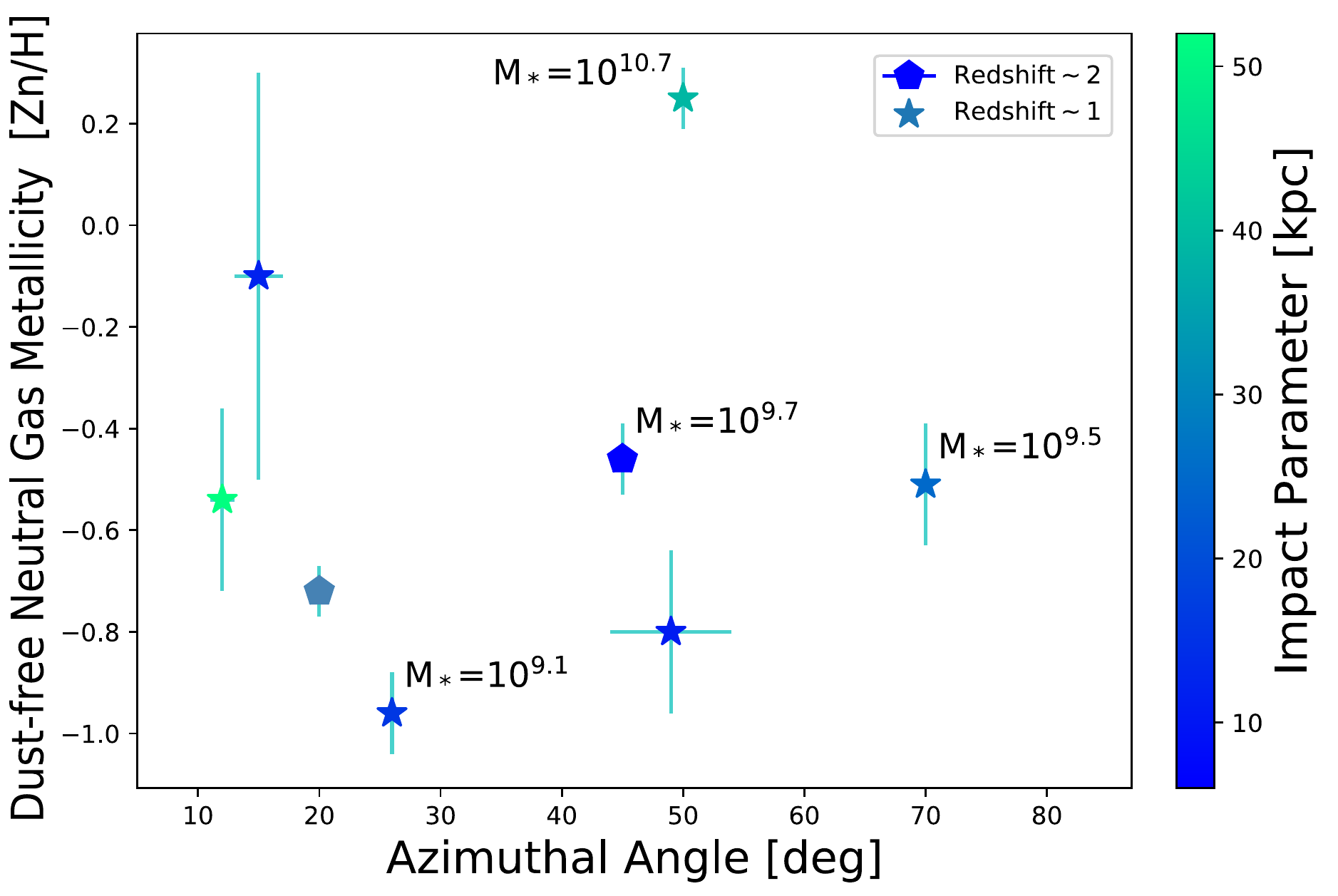}
    \caption{Observed relation between CGM metallicity and azimuthal angle. We show exclusively dust-depletion corrected estimates traced by [Zn/H] which are based on neutral absorbers selected solely on H\,I \citep{peroux2011, peroux2012, peroux2017, peroux2019, bouche2013, bouche2016}. At present, the sample size is too small and too heterogeneous, in stellar mass, redshift, and impact parameter, to compare to the simulations or to uncover any clear observational trend with azimuthal angle. However, a larger more homogeneous sample of galaxy-quasar absorber pairs could reveal the trend of metallicity with azimuthal angle, opening a diagnostic on inflowing versus outflowing gas.}
    \label{fig:obs_metallicity}
\end{figure}

The primary observational method for studying CGM gas is through absorption line spectroscopy, and the metallicity of absorbing gas offers clues as to its origin. Specifically, observations indicate a wide range of CGM metallicities \citep[spanning $>$2 dex;][]{lehner2013, fumagalli2016a, wotta2016, prochaska2017, zahedy2019}, presumably tracing both accretion and outflows. 

Lyman limit systems (LLSs, defined as systems optically thick at the Lyman limit with $16.2 < \log{(N_{\rm HI}/\rm{cm}^2)} < 17.0$) trace overdense structures in the haloes of galaxies \citep{ribaudo2011, tripp2011, fumagalli2011}. Intriguingly, \cite{wotta2016} and \cite{lehner2018} find that the metallicity distribution of LLSs is bimodal at $z < 1$, with a minimum at one tenth solar ($\log Z/\rm{Z}_\odot = -1.0$). The low-metallicity systems likely imply that the CGM of $z < 1$ star-forming galaxies is host to a substantial mass of cool, dense, metal-poor gas \citep{wotta2019}. The higher metallicity, \mbox{$Z > 0.1 \rm{Z}_\odot$} LLSs presumably arise in outflows, recycling winds, and tidally-stripped gas around galaxies. 

Therefore, studies of individual metal-line-selected absorbers have found tentative evidence for accretion and feedback signatures in the CGM. However, it has not yet been demonstrated that cosmological simulations can reproduce this bimodality \citep{hafen2017, rahmati2018, lehner2019}. Furthermore, it is unclear whether this bimodality is also observed at higher redshifts \citep{quiret2016, lehner2016}, and the overall interpretation is complicated by the metal-line selection of absorbers, potentially limiting detections of accretion, as well as the substantial ionisation corrections required \citep{fumagalli2016a}. 

To date, measurements of both the absorber metallicity and the orientation of the galaxy disk with respect to the quasar sightline are limited to only a few cases. With the benefit of a strong H\,I-selected quasar absorber sample, \cite{peroux2016} connected metallicity with azimuthal angle, although the small dataset indicated no sign of correlation. \cite{kacprzak2019b} recently expanded the samples to lower column densities but find a similar lack of trend. These works focused on the \textit{difference} between the galaxy metallicity, based on emission from H\,II regions, and the neutral gas metallicity, from absorption, in order to control for the stellar mass trend. The direct comparison of absorption gas metallicity with azimuthal angle, however, also shows no strong correlation \citep{pointon2019}. We note that these latter samples contain systems with column densities in the range $13.8 < \log{(N_{\rm HI}/\rm{cm}^2)} < 19.9$ and depend on non-negligible ionisation corrections. By restricting to mostly neutral quasar absorbers with $\log N_{\rm HI} > 19.5$\,cm$^{-2}$ one can largely circumvent these complications. 

An additional issue is the, a priori unknown, level of metal depletion onto dust grains. Dust is not yet commonly modeled in cosmological simulations, and is not included in either EAGLE or TNG50. However, dust grains are thought to play a central role in catalyzing the formation of the molecules essential to star formation and hence metal production, and there is a tight relation between metallicity and dust content, as seen in the evolution of the dust-to-gas ratio with metallicity \citep{jenkins2009, decia2016, peroux2020}. Hence, we expect the metallicity trend reported here to also appear versus dust or molecular content. In addition, gas-phase metals can be trapped in dust. As a result, if metallicity estimates are not corrected for dust-depletion the corresponding values may be biased \citep{kacprzak2019b,pointon2019}. 

To give a current view of the available data, we therefore show dust-depletion corrected [Zn/H] gas metallicity estimates based on dominantly neutral absorbers purely selected on H\,I. Figure~\ref{fig:obs_metallicity} displays the small sample of data currently available \citep[taken from][]{peroux2011, peroux2012, peroux2017, peroux2019, bouche2013, bouche2016}. Eight data points are available, and half have stellar masses, which range from $10^{9.1}$\,\msun\, to $10^{10.7}$\,\msun\, (indicated by text labels). The data also span a wide redshift range, where we separate $z \sim 1$ (star symbols) and $z \sim 2$ (pentagon symbols), as well as impact parameter, spanning $b = 6$\,pkpc to $b = 52$\,pkpc as indicated by the color bar.

Overall, the normalization of the metallicity values are characteristically higher than most of the simulation correlations explored in Section~\ref{sec:parameters}. The lack of low metallicity measurements in the data does not reflect a detection limit of the observations, which are typically sensitive to metallicity down to $\log{(Z/\rm{Z}_\odot)} = -2.5$ at $z < 1.5$ and as low as $\log{(Z/\rm{Z}_\odot)} = -3.5$ at higher redshifts \citep{peroux2020}. The high metallicities found by the observations are in part related to the low impact parameters probed in the range $b = 6$\,pkpc to $b = 52$\,pkpc. While the data span $-1.0 \lesssim \log{(Z_{\rm gas}/\rm{Z}_\odot)} \lesssim 0.2$ we have found in TNG50 that the highest of these metallicites are only likely to be found at the smallest impact parameters, $b < 25$\,pkpc, and around more massive galaxies, $M_\star \gtrsim 10^{10}$\,\msun\, (Figure~\ref{fig:Z_vs_angle_dependencies}, upper panels). Still, the gas more likely to be cosmic inflows has not been clearly detected, perhaps simply as a consequence of the different geometries, covering fractions, and available statistics.

Most importantly, it is clear that no strong trend of metallicity versus azimuthal angle of any kind can be inferred with the current data. As discussed earlier, we imagine that this is primarily because of the heterogeneous mix of the sample, and the broad range of impact parameters, stellar masses and redshifts in the dataset. Indeed, we would not expect to see the predicted trend, highlighting the current limits of our empirical constraints \citep{peroux2016, pointon2019, kacprzak2019b} and the utility of significantly larger data samples. We stress that an accurate measurement of the metallicity of the gas requires simultaneous knowledge of the hydrogen column density and an elemental abundance with minimal depletion onto dust grains. 

An empirical verification of the trend of increasing metallicity with azimuthal angle, as found in the EAGLE and TNG50 simulations and shown in Figure~\ref{fig:Z_vs_angle}, would lend strong support to key aspects of physical models of galaxy formation and evolution. For example, on numerical issues such as the treatment of sub-grid metal diffusion \citep{shen10,hafen2019}, as well as physical aspects, such as the expected mass loading factors from supernovae type II driven winds as a function of local gas properties \citep{schaye2015,pillepich18a}. Better measurements will also indirectly help constrain stellar population yields \citep[e.g.][]{kobayashi20}, particularly as more complete treatments of the relevant chemistry becomes common in cosmological simulations \citep{ploeckinger20}. 

Many physical processes and mechanisms contribute to the physical properties of gas in the circumgalactic medium, and mapping the metallicity throughout the halos of galaxies offers a key diagnostic of the cosmic baryon cycle.


\section{Conclusions} \label{sec:conclusions}

In this work we have used two recent cosmological galaxy simulations, TNG50 \citep{nelson19b,pillepich19} and EAGLE \citep{schaye2015,crain2015}, to explore the spatial distribution of gas in the multi-phase, circumgalactic medium (CGM) around galaxies. We focused on the dependence of gas mass flow rates and gas metallicity as a function of azimuthal angle with respect to the central galaxy as signposts capable of distinguishing inflow versus outflow. Our main results are:

\begin{itemize}
    \item Gas mass flow rates show clear variation with azimuthal angle, and are higher along directions aligned with both the major and minor axes of galaxies. High inflow rates preferentially trace accreting gas along the major axis, while high outflow rates more likely trace galactic winds along the minor axis. Results from TNG50 and EAGLE show a remarkable agreement of the predicted signal. The angular modulation of $\dot{M}$ is clear around the minor and major axes, where it becomes broadly compatible with observational data (Figure~\ref{fig:tng50_schematic} and Figure~\ref{fig:mass_flow_rate}).
    
    \item CGM gas metallicity correlates with radial velocity, demonstrating the connection between gas metallicity and flow direction. Outflows are found to have preferentially higher metallicity than inflows, making gas-phase metallicity a valuable indicator of the physical properties of the CGM (Figure~\ref{fig:Z_vs_velocity}). 

    \item The metallicity of gas in the CGM is a strong function of azimuthal angle for our fiducial choice of \mbox{$M_\star = 10^{9.5 \pm 0.1}$\,\msun at $z=0.5$ and impact parameter $b = 100 \pm 10$\,pkpc} (Figure~\ref{fig:Z_vs_angle}). The EAGLE and TNG50 simulations both predict a consistent signal, with metallicity higher along the minor axis than the major axis, reflecting the higher enrichment of galactic outflows versus cosmic gas accretion.

    \item The normalization of CGM metallicity varies strongly with impact parameter, stellar mass and redshift, increasing for larger $M_\star$, smaller $b$, and towards $z=0$. However, the correlation of metallicity with azimuthal angle is robust against different parameter choices, across a broad mass range of $8.5 < \log{(M_\star / \rm{M}_\odot)} < 10.5$, for all redshifts $z<1$, and particularly for $b \gtrsim 100$\,pkpc. This angular signal also remains when restricting to sightlines which intersect high neutral gas column densities, albeit with lower significance for the highest \nhi\ threshold (Figure~\ref{fig:Z_vs_angle_dependencies}).

    \item The data available to constrain this relation between CGM metallicity and azimuthal angle, based on [Zn/H] dust-free estimates, is currently too limited and too heterogeneous to (in)validate the simulations' outcome (Figure~\ref{fig:obs_metallicity}).
\end{itemize}

We have provided quantitative predictions for a CGM observable, focused on our ability to interpret forthcoming data from future large survey programs. Our results demonstrate that the azimuthal angle is a good discriminant to identify the gas flow direction. While fundamental, observational measurements of gas mass flow rates remain challenging because of the time derivative nature of the quantity, the numerous critical assumptions and the model-dependent approach. Our findings indicate that a statistically significant, controlled sample of galaxy-quasar absorber pairs should reveal a trend of metallicity with azimuthal angle that represents signatures of inflowing and outflowing gas.

The increasing availability of wide, high thoroughput and 3D integral-field unit spectrographs allows for studies of absorption fields which will enable statistically robust comparisons with these theoretical expectations. These include programs on VLT/MUSE (and Keck/KCWI), including MEGAFLOW, MUSEQuBES, MUSE-ALMA Halos, MAGG, and similar surveys \citep{zabl2019, muzahid2020, hamanowicz2020, lofthouse2020}. Our results reveal that a large sample of galaxy-absorber pairs with robust metallicity measurements (including mostly neutral gas traced by higher \nhi\ and reliable dust-depletion correction) with impact parameters up to b=200kpc at z$<$1 are best to uncover the trend of metallicity with azimuthal angle. Additional measures of the stellar masses of the galaxies are however essential. IFU-based kinematics of associated galaxies are also needed in order to determine their orientation (azimuthal angle) with respect to absorbing gas and thus constrain the relationship between galaxies and gas flows through their circumgalactic media.

Ultimately, observational signatures of both cosmic gas inflows and galactic outflows are important probes of how matter, energy, and metals move into and out of galaxies, and characterizing these flows provides unique insight into the processes of feeding and feedback, two crucial components of galaxy formation. 

\section*{Data Availability}

Data directly related to this publication and its figures is available on request from the corresponding author. The IllustrisTNG simulations themselves are publicly available and accessible at \url{www.tng-project.org/data}, where the TNG50 simulation will also be made public in the future. The EAGLE simulation is publicly available and accessible at \url{www.icc.dur.ac.uk/Eagle}.

\section*{Acknowledgements}

CP thanks the Alexander von Humboldt Foundation for the granting of a Bessel Research Award held at MPA where this work was initiated. FvdV acknowledges support from a Royal Society University Research Fellowship. FM acknowledges support through the Program "Rita Levi Montalcini" of the Italian MIUR. The TNG50 simulation was realized with compute time granted by the Gauss Centre for Supercomputing (GCS) under GCS Large-Scale Projects GCS-DWAR on the GCS share of the supercomputer Hazel Hen at the High Performance Computing Center Stuttgart (HLRS). GCS is the alliance of the three national supercomputing centres HLRS (Universit{\"a}t Stuttgart), JSC (Forschungszentrum J{\"u}lich), and LRZ (Bayerische Akademie der Wissenschaften), funded by the German Federal Ministry of Education and Research (BMBF) and the German State Ministries for Research of Baden-W{\"u}rttemberg (MWK), Bayern (StMWFK) and Nordrhein-Westfalen (MIWF). Additional simulations and analysis were carried out on the supercomputers at the Max Planck Computing and Data Facility (MPCDF).

\bibliographystyle{mnras}
\bibliography{Biblio}

\begin{thebibliography}{}
\makeatletter
\relax
\def\mn@urlcharsother{\let\do\@makeother \do\$\do\&\do\#\do\^\do\_\do\%\do\~}
\def\mn@doi{\begingroup\mn@urlcharsother \@ifnextchar [ {\mn@doi@}
  {\mn@doi@[]}}
\def\mn@doi@[#1]#2{\def\@tempa{#1}\ifx\@tempa\@empty \href
  {http://dx.doi.org/#2} {doi:#2}\else \href {http://dx.doi.org/#2} {#1}\fi
  \endgroup}
\def\mn@eprint#1#2{\mn@eprint@#1:#2::\@nil}
\def\mn@eprint@arXiv#1{\href {http://arxiv.org/abs/#1} {{\tt arXiv:#1}}}
\def\mn@eprint@dblp#1{\href {http://dblp.uni-trier.de/rec/bibtex/#1.xml}
  {dblp:#1}}
\def\mn@eprint@#1:#2:#3:#4\@nil{\def\@tempa {#1}\def\@tempb {#2}\def\@tempc
  {#3}\ifx \@tempc \@empty \let \@tempc \@tempb \let \@tempb \@tempa \fi \ifx
  \@tempb \@empty \def\@tempb {arXiv}\fi \@ifundefined
  {mn@eprint@\@tempb}{\@tempb:\@tempc}{\expandafter \expandafter \csname
  mn@eprint@\@tempb\endcsname \expandafter{\@tempc}}}

\bibitem[\protect\citeauthoryear{Adelberger, Steidel, Shapley  \&
  Pettini}{Adelberger et~al.}{2003}]{adelberger2003}
Adelberger K.,  Steidel C.,  Shapley A.,   Pettini M.,  2003, ApJ, 584, 45

\bibitem[\protect\citeauthoryear{Adelberger, Shapley, Steidel, Pettini, Erb  \&
  Reddy}{Adelberger et~al.}{2005}]{adelberger2005}
Adelberger K.,  Shapley A.,  Steidel C.,  Pettini M.,  Erb D.,   Reddy N.,
  2005, ApJ, 629, 636

\bibitem[\protect\citeauthoryear{{Aguirre}, {Hernquist}, {Schaye}, {Weinberg},
  {Katz}  \& {Gardner}}{{Aguirre} et~al.}{2001}]{aguirre2001}
{Aguirre} A.,  {Hernquist} L.,  {Schaye} J.,  {Weinberg} D.~H.,  {Katz} N.,
  {Gardner} J.,  2001, \mn@doi [ApJ] {10.1086/323070}, \href
  {http://adsabs.harvard.edu/abs/2001ApJ...560..599A} {560, 599}

\bibitem[\protect\citeauthoryear{{Angl{\'e}s-Alc{\'a}zar}, {Dav{\'e}},
  {{\"O}zel}  \& {Oppenheimer}}{{Angl{\'e}s-Alc{\'a}zar}
  et~al.}{2014}]{angles2014}
{Angl{\'e}s-Alc{\'a}zar} D.,  {Dav{\'e}} R.,  {{\"O}zel} F.,   {Oppenheimer}
  B.~D.,  2014, \mn@doi [\apj] {10.1088/0004-637X/782/2/84}, \href
  {https://ui.adsabs.harvard.edu/abs/2014ApJ...782...84A} {782, 84}

\bibitem[\protect\citeauthoryear{{Angl{\'e}s-Alc{\'a}zar},
  {Faucher-Gigu{\`e}re}, {Kere{\v{s}}}, {Hopkins}, {Quataert}  \&
  {Murray}}{{Angl{\'e}s-Alc{\'a}zar} et~al.}{2017}]{angles2017}
{Angl{\'e}s-Alc{\'a}zar} D.,  {Faucher-Gigu{\`e}re} C.-A.,  {Kere{\v{s}}} D.,
  {Hopkins} P.~F.,  {Quataert} E.,   {Murray} N.,  2017, \mn@doi [\mnras]
  {10.1093/mnras/stx1517}, \href
  {https://ui.adsabs.harvard.edu/abs/2017MNRAS.470.4698A} {470, 4698}

\bibitem[\protect\citeauthoryear{{Augustin} et~al.,}{{Augustin}
  et~al.}{2018}]{augustin2018}
{Augustin} R.,  et~al., 2018, \mn@doi [\mnras] {10.1093/mnras/sty1287}, \href
  {https://ui.adsabs.harvard.edu/abs/2018MNRAS.478.3120A} {478, 3120}

\bibitem[\protect\citeauthoryear{{Augustin} et~al.,}{{Augustin}
  et~al.}{2019}]{augustin2019}
{Augustin} R.,  et~al., 2019, \mn@doi [\mnras] {10.1093/mnras/stz2238}, \href
  {https://ui.adsabs.harvard.edu/abs/2019MNRAS.489.2417A} {489, 2417}

\bibitem[\protect\citeauthoryear{{Bertone}, {Schaye}, {Dalla Vecchia}, {Booth},
  {Theuns}  \& {Wiersma}}{{Bertone} et~al.}{2010a}]{bertone2010a}
{Bertone} S.,  {Schaye} J.,  {Dalla Vecchia} C.,  {Booth} C.~M.,  {Theuns} T.,
   {Wiersma} R. P.~C.,  2010a, \mn@doi [\mnras]
  {10.1111/j.1365-2966.2010.16932.x}, \href
  {https://ui.adsabs.harvard.edu/abs/2010MNRAS.407..544B} {407, 544}

\bibitem[\protect\citeauthoryear{{Bertone}, {Schaye}, {Booth}, {Dalla Vecchia},
  {Theuns}  \& {Wiersma}}{{Bertone} et~al.}{2010b}]{bertone2010b}
{Bertone} S.,  {Schaye} J.,  {Booth} C.~M.,  {Dalla Vecchia} C.,  {Theuns} T.,
   {Wiersma} R. P.~C.,  2010b, \mn@doi [\mnras]
  {10.1111/j.1365-2966.2010.17188.x}, \href
  {https://ui.adsabs.harvard.edu/abs/2010MNRAS.408.1120B} {408, 1120}

\bibitem[\protect\citeauthoryear{Bordoloi et~al.,}{Bordoloi
  et~al.}{2011}]{bordoloi2011}
Bordoloi R.,  et~al., 2011, ApJ, 743, 10

\bibitem[\protect\citeauthoryear{Bouch\'e, Murphy, P\'eroux, Davies,
  Eisenhauer, Forster-Schreiber  \& Tacconi}{Bouch\'e
  et~al.}{2007a}]{bouche2007a}
Bouch\'e N. N.,  Murphy M.,  P\'eroux C.,  Davies R.,  Eisenhauer F.,
  Forster-Schreiber N.,   Tacconi L.,  2007a, ApJ, 669, L5

\bibitem[\protect\citeauthoryear{{Bouch{\'e}}, {Murphy}, {P{\'e}roux},
  {Davies}, {Eisenhauer}, {F{\"o}rster Schreiber}  \& {Tacconi}}{{Bouch{\'e}}
  et~al.}{2007b}]{bouche2007}
{Bouch{\'e}} N.,  {Murphy} M.~T.,  {P{\'e}roux} C.,  {Davies} R.,  {Eisenhauer}
  F.,  {F{\"o}rster Schreiber} N.~M.,   {Tacconi} L.,  2007b, \mn@doi [\apjl]
  {10.1086/523594}, \href
  {https://ui.adsabs.harvard.edu/abs/2007ApJ...669L...5B} {669, L5}

\bibitem[\protect\citeauthoryear{Bouch\'e et~al.,}{Bouch\'e
  et~al.}{2012}]{bouche2012b}
Bouch\'e N.,  et~al., 2012, MNRAS, 419, 2

\bibitem[\protect\citeauthoryear{Bouch\'e, Murphy, P\'eroux, Contini, Martin
  \& Dessauges-Zavadsky}{Bouch\'e et~al.}{2013}]{bouche2013}
Bouch\'e N.,  Murphy M.,  P\'eroux C.,  Contini T.,  Martin C.,
  Dessauges-Zavadsky M.,  2013, Sci, 341, 50

\bibitem[\protect\citeauthoryear{{Bouch{\'e}} et~al.,}{{Bouch{\'e}}
  et~al.}{2016}]{bouche2016}
{Bouch{\'e}} N.,  et~al., 2016, \mn@doi [ApJ] {10.3847/0004-637X/820/2/121},
  \href {http://adsabs.harvard.edu/abs/2016ApJ...820..121B} {820, 121}

\bibitem[\protect\citeauthoryear{{Bower}, {Schaye}, {Frenk}, {Theuns},
  {Schaller}, {Crain}  \& {McAlpine}}{{Bower} et~al.}{2017}]{bower17}
{Bower} R.~G.,  {Schaye} J.,  {Frenk} C.~S.,  {Theuns} T.,  {Schaller} M.,
  {Crain} R.~A.,   {McAlpine} S.,  2017, \mn@doi [\mnras]
  {10.1093/mnras/stw2735}, \href
  {https://ui.adsabs.harvard.edu/abs/2017MNRAS.465...32B} {465, 32}

\bibitem[\protect\citeauthoryear{{Brook} et~al.,}{{Brook}
  et~al.}{2011}]{brook2011}
{Brook} C.~B.,  et~al., 2011, \mn@doi [MNRAS]
  {10.1111/j.1365-2966.2011.18545.x}, \href
  {http://adsabs.harvard.edu/abs/2011MNRAS.415.1051B} {415, 1051}

\bibitem[\protect\citeauthoryear{{Cen} \& {Chisari}}{{Cen} \&
  {Chisari}}{2011}]{cen2011}
{Cen} R.,  {Chisari} N.~E.,  2011, \mn@doi [\apj] {10.1088/0004-637X/731/1/11},
  \href {https://ui.adsabs.harvard.edu/abs/2011ApJ...731...11C} {731, 11}

\bibitem[\protect\citeauthoryear{{Cen}, {Nagamine}  \& {Ostriker}}{{Cen}
  et~al.}{2005}]{cen2005}
{Cen} R.,  {Nagamine} K.,   {Ostriker} J.~P.,  2005, \mn@doi [\apj]
  {10.1086/497353}, \href
  {https://ui.adsabs.harvard.edu/abs/2005ApJ...635...86C} {635, 86}

\bibitem[\protect\citeauthoryear{{Chen} et~al.,}{{Chen} et~al.}{2017}]{chen17}
{Chen} S.-F.~S.,  et~al., 2017, \mn@doi [\apj] {10.3847/1538-4357/aa9707},
  \href {http://adsabs.harvard.edu/abs/2017ApJ...850..188C} {850, 188}

\bibitem[\protect\citeauthoryear{{Corlies}, {Peeples}, {Tumlinson}, {O'Shea},
  {Lehner}, {Howk}, {O'Meara}  \& {Smith}}{{Corlies} et~al.}{2020}]{corlies20}
{Corlies} L.,  {Peeples} M.~S.,  {Tumlinson} J.,  {O'Shea} B.~W.,  {Lehner} N.,
   {Howk} J.~C.,  {O'Meara} J.~M.,   {Smith} B.~D.,  2020, \mn@doi [\apj]
  {10.3847/1538-4357/ab9310}, \href
  {https://ui.adsabs.harvard.edu/abs/2020ApJ...896..125C} {896, 125}

\bibitem[\protect\citeauthoryear{{Crain} et~al.,}{{Crain}
  et~al.}{2015}]{crain2015}
{Crain} R.~A.,  et~al., 2015, \mn@doi [\mnras] {10.1093/mnras/stv725}, \href
  {https://ui.adsabs.harvard.edu/abs/2015MNRAS.450.1937C} {450, 1937}

\bibitem[\protect\citeauthoryear{{Dalla Vecchia} \& {Schaye}}{{Dalla Vecchia}
  \& {Schaye}}{2012}]{dellavecchia2012}
{Dalla Vecchia} C.,  {Schaye} J.,  2012, \mn@doi [\mnras]
  {10.1111/j.1365-2966.2012.21704.x}, \href
  {https://ui.adsabs.harvard.edu/abs/2012MNRAS.426..140D} {426, 140}

\bibitem[\protect\citeauthoryear{{Dav{\'e}}, {Thompson}  \&
  {Hopkins}}{{Dav{\'e}} et~al.}{2016}]{dave16}
{Dav{\'e}} R.,  {Thompson} R.,   {Hopkins} P.~F.,  2016, \mn@doi [\mnras]
  {10.1093/mnras/stw1862}, \href
  {http://adsabs.harvard.edu/abs/2016MNRAS.462.3265D} {462, 3265}

\bibitem[\protect\citeauthoryear{{Davis}, {Efstathiou}, {Frenk}  \&
  {White}}{{Davis} et~al.}{1985}]{davis85}
{Davis} M.,  {Efstathiou} G.,  {Frenk} C.~S.,   {White} S.~D.~M.,  1985,
  \mn@doi [\apj] {10.1086/163168}, \href
  {http://adsabs.harvard.edu/abs/1985ApJ...292..371D} {292, 371}

\bibitem[\protect\citeauthoryear{{De Cia}, {Ledoux}, {Mattsson}, {Petitjean},
  {Srianand}, {Gavignaud}  \& {Jenkins}}{{De Cia} et~al.}{2016}]{decia2016}
{De Cia} A.,  {Ledoux} C.,  {Mattsson} L.,  {Petitjean} P.,  {Srianand} R.,
  {Gavignaud} I.,   {Jenkins} E.~B.,  2016, \mn@doi [\aap]
  {10.1051/0004-6361/201527895}, \href
  {https://ui.adsabs.harvard.edu/#abs/2016A&A...596A..97D} {596, A97}

\bibitem[\protect\citeauthoryear{{De Cia}, {Ledoux}, {Petitjean}  \&
  {Savaglio}}{{De Cia} et~al.}{2018}]{decia2018}
{De Cia} A.,  {Ledoux} C.,  {Petitjean} P.,   {Savaglio} S.,  2018, \mn@doi
  [\aap] {10.1051/0004-6361/201731970}, \href
  {http://adsabs.harvard.edu/abs/2018A%26A...611A..76D} {611, A76}

\bibitem[\protect\citeauthoryear{{DeFelippis}, {Genel}, {Bryan}, {Nelson},
  {Pillepich}  \& {Hernquist}}{{DeFelippis} et~al.}{2020}]{defelippis20}
{DeFelippis} D.,  {Genel} S.,  {Bryan} G.~L.,  {Nelson} D.,  {Pillepich} A.,
  {Hernquist} L.,  2020, \mn@doi [\apj] {10.3847/1538-4357/ab8a4a}, \href
  {https://ui.adsabs.harvard.edu/abs/2020ApJ...895...17D} {895, 17}

\bibitem[\protect\citeauthoryear{{Dekel} et~al.,}{{Dekel}
  et~al.}{2009}]{dekel09}
{Dekel} A.,  et~al., 2009, \mn@doi [\nat] {10.1038/nature07648}, \href
  {http://adsabs.harvard.edu/abs/2009Natur.457..451D} {457, 451}

\bibitem[\protect\citeauthoryear{{Di Matteo}, {Springel}  \& {Hernquist}}{{Di
  Matteo} et~al.}{2005}]{dimatteo05}
{Di Matteo} T.,  {Springel} V.,   {Hernquist} L.,  2005, \mn@doi [\nat]
  {10.1038/nature03335}, \href
  {https://ui.adsabs.harvard.edu/abs/2005Natur.433..604D} {433, 604}

\bibitem[\protect\citeauthoryear{{Donnari} et~al.,}{{Donnari}
  et~al.}{2019}]{donnari19}
{Donnari} M.,  et~al., 2019, \mn@doi [\mnras] {10.1093/mnras/stz712}, \href
  {https://ui.adsabs.harvard.edu/abs/2019MNRAS.485.4817D} {485, 4817}

\bibitem[\protect\citeauthoryear{{Dubois} et~al.,}{{Dubois}
  et~al.}{2014}]{dubois14}
{Dubois} Y.,  et~al., 2014, \mn@doi [\mnras] {10.1093/mnras/stu1227}, \href
  {http://adsabs.harvard.edu/abs/2014MNRAS.444.1453D} {444, 1453}

\bibitem[\protect\citeauthoryear{{Faucher-Gigu{\`e}re}, {Lidz}, {Zaldarriaga}
  \& {Hernquist}}{{Faucher-Gigu{\`e}re} et~al.}{2009}]{faucher09}
{Faucher-Gigu{\`e}re} C.-A.,  {Lidz} A.,  {Zaldarriaga} M.,   {Hernquist} L.,
  2009, \mn@doi [\apj] {10.1088/0004-637X/703/2/1416}, \href
  {https://ui.adsabs.harvard.edu/abs/2009ApJ...703.1416F} {703, 1416}

\bibitem[\protect\citeauthoryear{{Faucher-Gigu{\`e}re}, {Feldmann}, {Quataert},
  {Kere{\v{s}}}, {Hopkins}  \& {Murray}}{{Faucher-Gigu{\`e}re}
  et~al.}{2016}]{fg16}
{Faucher-Gigu{\`e}re} C.-A.,  {Feldmann} R.,  {Quataert} E.,  {Kere{\v{s}}} D.,
   {Hopkins} P.~F.,   {Murray} N.,  2016, \mn@doi [\mnras]
  {10.1093/mnrasl/slw091}, \href
  {https://ui.adsabs.harvard.edu/abs/2016MNRAS.461L..32F} {461, L32}

\bibitem[\protect\citeauthoryear{{Frank} et~al.,}{{Frank}
  et~al.}{2012}]{frank2012}
{Frank} S.,  et~al., 2012, \mn@doi [\mnras] {10.1111/j.1365-2966.2011.20172.x},
  \href {https://ui.adsabs.harvard.edu/abs/2012MNRAS.420.1731F} {420, 1731}

\bibitem[\protect\citeauthoryear{{Fraternali} \& {Binney}}{{Fraternali} \&
  {Binney}}{2008}]{fraternali2008}
{Fraternali} F.,  {Binney} J.~J.,  2008, \mn@doi [\mnras]
  {10.1111/j.1365-2966.2008.13071.x}, \href
  {https://ui.adsabs.harvard.edu/abs/2008MNRAS.386..935F} {386, 935}

\bibitem[\protect\citeauthoryear{{Fumagalli}, {O'Meara}  \&
  {Prochaska}}{{Fumagalli} et~al.}{2011}]{fumagalli2011}
{Fumagalli} M.,  {O'Meara} J.~M.,   {Prochaska} J.~X.,  2011, \mn@doi [Science]
  {10.1126/science.1213581}, \href
  {https://ui.adsabs.harvard.edu/#abs/2011Sci...334.1245F} {334, 1245}

\bibitem[\protect\citeauthoryear{{Fumagalli}, {Cantalupo}, {Dekel}, {Morris},
  {O'Meara}, {Prochaska}  \& {Theuns}}{{Fumagalli}
  et~al.}{2016}]{fumagalli2016a}
{Fumagalli} M.,  {Cantalupo} S.,  {Dekel} A.,  {Morris} S.~L.,  {O'Meara}
  J.~M.,  {Prochaska} J.~X.,   {Theuns} T.,  2016, \mn@doi [MNRAS]
  {10.1093/mnras/stw1782}, \href
  {http://adsabs.harvard.edu/abs/2016MNRAS.462.1978F} {462, 1978}

\bibitem[\protect\citeauthoryear{{Grand} et~al.,}{{Grand}
  et~al.}{2019}]{grand19}
{Grand} R. J.~J.,  et~al., 2019, \mn@doi [\mnras] {10.1093/mnras/stz2928},
  \href {https://ui.adsabs.harvard.edu/abs/2019MNRAS.490.4786G} {490, 4786}

\bibitem[\protect\citeauthoryear{{Haardt} \& {Madau}}{{Haardt} \&
  {Madau}}{2001}]{haardt2001}
{Haardt} F.,  {Madau} P.,  2001, in Clusters of Galaxies and the High Redshift
  Universe Observed in X-rays. p.~64 (\mn@eprint {arXiv} {astro-ph/0106018})

\bibitem[\protect\citeauthoryear{{Hafen} et~al.,}{{Hafen}
  et~al.}{2017}]{hafen2017}
{Hafen} Z.,  et~al., 2017, \mn@doi [\mnras] {10.1093/mnras/stx952}, \href
  {https://ui.adsabs.harvard.edu/abs/2017MNRAS.469.2292H} {469, 2292}

\bibitem[\protect\citeauthoryear{{Hafen} et~al.,}{{Hafen}
  et~al.}{2019}]{hafen2019}
{Hafen} Z.,  et~al., 2019, \mn@doi [\mnras] {10.1093/mnras/stz1773}, \href
  {https://ui.adsabs.harvard.edu/abs/2019MNRAS.488.1248H} {488, 1248}

\bibitem[\protect\citeauthoryear{{Hafen} et~al.,}{{Hafen}
  et~al.}{2020}]{hafen2020}
{Hafen} Z.,  et~al., 2020, \mn@doi [\mnras] {10.1093/mnras/staa902}, \href
  {https://ui.adsabs.harvard.edu/abs/2020MNRAS.494.3581H} {494, 3581}

\bibitem[\protect\citeauthoryear{{Hamanowicz}, {Peroux}, {Zwaan}  \&
  {Rahmani}}{{Hamanowicz} et~al.}{2020}]{hamanowicz2020}
{Hamanowicz} A.,  {Peroux} C.,  {Zwaan} M.~A.,   {Rahmani} H. e.~a.,  2020,
  MNRAS, Submitted

\bibitem[\protect\citeauthoryear{{Ho}, {Martin}  \& {Turner}}{{Ho}
  et~al.}{2019}]{ho19}
{Ho} S.~H.,  {Martin} C.~L.,   {Turner} M.~L.,  2019, \mn@doi [\apj]
  {10.3847/1538-4357/ab0ec2}, \href
  {https://ui.adsabs.harvard.edu/abs/2019ApJ...875...54H} {875, 54}

\bibitem[\protect\citeauthoryear{{Hopkins}, {Kere{\v{s}}}, {O{\~n}orbe},
  {Faucher-Gigu{\`e}re}, {Quataert}, {Murray}  \& {Bullock}}{{Hopkins}
  et~al.}{2014}]{hopkins2014}
{Hopkins} P.~F.,  {Kere{\v{s}}} D.,  {O{\~n}orbe} J.,  {Faucher-Gigu{\`e}re}
  C.-A.,  {Quataert} E.,  {Murray} N.,   {Bullock} J.~S.,  2014, \mn@doi
  [\mnras] {10.1093/mnras/stu1738}, \href
  {https://ui.adsabs.harvard.edu/abs/2014MNRAS.445..581H} {445, 581}

\bibitem[\protect\citeauthoryear{{Hummels} et~al.,}{{Hummels}
  et~al.}{2019}]{hummels19}
{Hummels} C.~B.,  et~al., 2019, \mn@doi [\apj] {10.3847/1538-4357/ab378f},
  \href {https://ui.adsabs.harvard.edu/abs/2019ApJ...882..156H} {882, 156}

\bibitem[\protect\citeauthoryear{{Jenkins}}{{Jenkins}}{2009}]{jenkins2009}
{Jenkins} E.~B.,  2009, \mn@doi [\apj] {10.1088/0004-637X/700/2/1299}, \href
  {http://adsabs.harvard.edu/abs/2009ApJ...700.1299J} {700, 1299}

\bibitem[\protect\citeauthoryear{{Kacprzak}, {Churchill}  \&
  {Nielsen}}{{Kacprzak} et~al.}{2012}]{kacprzak2012a}
{Kacprzak} G.~G.,  {Churchill} C.~W.,   {Nielsen} N.~M.,  2012, \mn@doi [\apjl]
  {10.1088/2041-8205/760/1/L7}, \href
  {https://ui.adsabs.harvard.edu/abs/2012ApJ...760L...7K} {760, L7}

\bibitem[\protect\citeauthoryear{{Kacprzak} et~al.,}{{Kacprzak}
  et~al.}{2014}]{kacprzak2014}
{Kacprzak} G.~G.,  et~al., 2014, \mn@doi [ApJL] {10.1088/2041-8205/792/1/L12},
  \href {http://adsabs.harvard.edu/abs/2014ApJ...792L..12K} {792, L12}

\bibitem[\protect\citeauthoryear{{Kacprzak}, {Churchill}, {Murphy}  \&
  {Cooke}}{{Kacprzak} et~al.}{2015a}]{kacprzak2015a}
{Kacprzak} G.~G.,  {Churchill} C.~W.,  {Murphy} M.~T.,   {Cooke} J.,  2015a,
  \mn@doi [\mnras] {10.1093/mnras/stu2324}, \href
  {https://ui.adsabs.harvard.edu/abs/2015MNRAS.446.2861K} {446, 2861}

\bibitem[\protect\citeauthoryear{{Kacprzak}, {Muzahid}, {Churchill}, {Nielsen}
  \& {Charlton}}{{Kacprzak} et~al.}{2015b}]{kacprzak2015c}
{Kacprzak} G.~G.,  {Muzahid} S.,  {Churchill} C.~W.,  {Nielsen} N.~M.,
  {Charlton} J.~C.,  2015b, \mn@doi [\apj] {10.1088/0004-637X/815/1/22}, \href
  {https://ui.adsabs.harvard.edu/abs/2015ApJ...815...22K} {815, 22}

\bibitem[\protect\citeauthoryear{{Kacprzak}, {Pointon}, {Nielsen}, {Churchill},
  {Muzahid}  \& {Charlton}}{{Kacprzak} et~al.}{2019}]{kacprzak2019b}
{Kacprzak} G.~G.,  {Pointon} S.~K.,  {Nielsen} N.~M.,  {Churchill} C.~W.,
  {Muzahid} S.,   {Charlton} J.~C.,  2019, \mn@doi [\apj]
  {10.3847/1538-4357/ab4c3c}, \href
  {https://ui.adsabs.harvard.edu/abs/2019ApJ...886...91K} {886, 91}

\bibitem[\protect\citeauthoryear{Kennicutt}{Kennicutt}{1998}]{kennicutt1998}
Kennicutt R.,  1998, ARAA, 36, 189

\bibitem[\protect\citeauthoryear{{Kere{\v s}}, {Katz}, {Weinberg}  \&
  {Dav{\'e}}}{{Kere{\v s}} et~al.}{2005}]{keres05}
{Kere{\v s}} D.,  {Katz} N.,  {Weinberg} D.~H.,   {Dav{\'e}} R.,  2005, \mn@doi
  [\mnras] {10.1111/j.1365-2966.2005.09451.x}, \href
  {http://adsabs.harvard.edu/abs/2005MNRAS.363....2K} {363, 2}

\bibitem[\protect\citeauthoryear{{Klitsch}, {P{\'e}roux}, {Zwaan}, {Smail},
  {Oteo}, {Biggs}, {Popping}  \& {Swinbank}}{{Klitsch}
  et~al.}{2018}]{klitsch2018}
{Klitsch} A.,  {P{\'e}roux} C.,  {Zwaan} M.~A.,  {Smail} I.,  {Oteo} I.,
  {Biggs} A.~D.,  {Popping} G.,   {Swinbank} A.~M.,  2018, \mn@doi [MNRAS]
  {10.1093/mnras/stx3184}, \href
  {http://adsabs.harvard.edu/abs/2018MNRAS.475..492K} {475, 492}

\bibitem[\protect\citeauthoryear{{Kobayashi}, {Leung}  \& {Nomoto}}{{Kobayashi}
  et~al.}{2020}]{kobayashi20}
{Kobayashi} C.,  {Leung} S.-C.,   {Nomoto} K.,  2020, \mn@doi [\apj]
  {10.3847/1538-4357/ab8e44}, \href
  {https://ui.adsabs.harvard.edu/abs/2020ApJ...895..138K} {895, 138}

\bibitem[\protect\citeauthoryear{{Lan}}{{Lan}}{2019}]{lan20}
{Lan} T.-W.,  2019, arXiv e-prints, \href
  {https://ui.adsabs.harvard.edu/abs/2019arXiv191101271L} {p. arXiv:1911.01271}

\bibitem[\protect\citeauthoryear{{Lehner} et~al.,}{{Lehner}
  et~al.}{2013}]{lehner2013}
{Lehner} N.,  et~al., 2013, \mn@doi [\apj] {10.1088/0004-637X/770/2/138}, \href
  {https://ui.adsabs.harvard.edu/#abs/2013ApJ...770..138L} {770, 138}

\bibitem[\protect\citeauthoryear{{Lehner}, {O'Meara}, {Howk}, {Prochaska}  \&
  {Fumagalli}}{{Lehner} et~al.}{2016}]{lehner2016}
{Lehner} N.,  {O'Meara} J.~M.,  {Howk} J.~C.,  {Prochaska} J.~X.,   {Fumagalli}
  M.,  2016, \mn@doi [\apj] {10.3847/1538-4357/833/2/283}, \href
  {https://ui.adsabs.harvard.edu/abs/2016ApJ...833..283L} {833, 283}

\bibitem[\protect\citeauthoryear{{Lehner}, {Wotta}, {Howk}, {O'Meara},
  {Oppenheimer}  \& {Cooksey}}{{Lehner} et~al.}{2018}]{lehner2018}
{Lehner} N.,  {Wotta} C.~B.,  {Howk} J.~C.,  {O'Meara} J.~M.,  {Oppenheimer}
  B.~D.,   {Cooksey} K.~L.,  2018, \mn@doi [\apj] {10.3847/1538-4357/aadd03},
  \href {https://ui.adsabs.harvard.edu/abs/2018ApJ...866...33L} {866, 33}

\bibitem[\protect\citeauthoryear{{Lehner}, {Wotta}, {Howk},
  {O{\textquoteright}Meara}, {Oppenheimer}  \& {Cooksey}}{{Lehner}
  et~al.}{2019}]{lehner2019}
{Lehner} N.,  {Wotta} C.~B.,  {Howk} J.~C.,  {O{\textquoteright}Meara} J.~M.,
  {Oppenheimer} B.~D.,   {Cooksey} K.~L.,  2019, \mn@doi [\apj]
  {10.3847/1538-4357/ab41fd}, \href
  {https://ui.adsabs.harvard.edu/abs/2019ApJ...887....5L} {887, 5}

\bibitem[\protect\citeauthoryear{{Lofthouse} et~al.,}{{Lofthouse}
  et~al.}{2020}]{lofthouse2020}
{Lofthouse} E.~K.,  et~al., 2020, \mn@doi [\mnras] {10.1093/mnras/stz3066},
  \href {https://ui.adsabs.harvard.edu/abs/2020MNRAS.491.2057L} {491, 2057}

\bibitem[\protect\citeauthoryear{{Mandelker}, {Nagai}, {Aung}, {Dekel},
  {Padnos}  \& {Birnboim}}{{Mandelker} et~al.}{2019}]{mandelker19}
{Mandelker} N.,  {Nagai} D.,  {Aung} H.,  {Dekel} A.,  {Padnos} D.,
  {Birnboim} Y.,  2019, \mn@doi [\mnras] {10.1093/mnras/stz012}, \href
  {https://ui.adsabs.harvard.edu/abs/2019MNRAS.484.1100M} {484, 1100}

\bibitem[\protect\citeauthoryear{{Marinacci} et~al.,}{{Marinacci}
  et~al.}{2018}]{marinacci18}
{Marinacci} F.,  et~al., 2018, \mn@doi [\mnras] {10.1093/mnras/sty2206}, \href
  {http://adsabs.harvard.edu/abs/2018MNRAS.480.5113M} {480, 5113}

\bibitem[\protect\citeauthoryear{{Martin} et~al.,}{{Martin}
  et~al.}{2019a}]{martin19}
{Martin} D.~C.,  et~al., 2019a, \mn@doi [Nature Astronomy]
  {10.1038/s41550-019-0791-2}, \href
  {https://ui.adsabs.harvard.edu/abs/2019NatAs...3..822M} {3, 822}

\bibitem[\protect\citeauthoryear{{Martin}, {Ho}, {Kacprzak}  \&
  {Churchill}}{{Martin} et~al.}{2019b}]{martin2019}
{Martin} C.~L.,  {Ho} S.~H.,  {Kacprzak} G.~G.,   {Churchill} C.~W.,  2019b,
  \mn@doi [\apj] {10.3847/1538-4357/ab18ac}, \href
  {https://ui.adsabs.harvard.edu/abs/2019ApJ...878...84M} {878, 84}

\bibitem[\protect\citeauthoryear{{McAlpine} et~al.,}{{McAlpine}
  et~al.}{2016}]{mcalpine2016}
{McAlpine} S.,  et~al., 2016, \mn@doi [Astronomy and Computing]
  {10.1016/j.ascom.2016.02.004}, \href
  {https://ui.adsabs.harvard.edu/abs/2016A&C....15...72M} {15, 72}

\bibitem[\protect\citeauthoryear{{Mitchell}, {Schaye}  \& {Bower}}{{Mitchell}
  et~al.}{2020a}]{mitchell20b}
{Mitchell} P.~D.,  {Schaye} J.,   {Bower} R.~G.,  2020a, arXiv e-prints, \href
  {https://ui.adsabs.harvard.edu/abs/2020arXiv200510262M} {p. arXiv:2005.10262}

\bibitem[\protect\citeauthoryear{{Mitchell}, {Schaye}, {Bower}  \&
  {Crain}}{{Mitchell} et~al.}{2020b}]{mitchell20a}
{Mitchell} P.~D.,  {Schaye} J.,  {Bower} R.~G.,   {Crain} R.~A.,  2020b,
  \mn@doi [\mnras] {10.1093/mnras/staa938}, \href
  {https://ui.adsabs.harvard.edu/abs/2020MNRAS.494.3971M} {494, 3971}

\bibitem[\protect\citeauthoryear{{Muratov} et~al.,}{{Muratov}
  et~al.}{2017}]{muratov17}
{Muratov} A.~L.,  et~al., 2017, \mn@doi [\mnras] {10.1093/mnras/stx667}, \href
  {http://adsabs.harvard.edu/abs/2017MNRAS.468.4170M} {468, 4170}

\bibitem[\protect\citeauthoryear{{Muzahid} et~al.,}{{Muzahid}
  et~al.}{2020}]{muzahid2020}
{Muzahid} S.,  et~al., 2020, \mn@doi [\mnras] {10.1093/mnras/staa1347}, \href
  {https://ui.adsabs.harvard.edu/abs/2020MNRAS.tmp.1463M} {}

\bibitem[\protect\citeauthoryear{{Naiman} et~al.,}{{Naiman}
  et~al.}{2018}]{naiman18}
{Naiman} J.~P.,  et~al., 2018, \mn@doi [\mnras] {10.1093/mnras/sty618}, \href
  {http://adsabs.harvard.edu/abs/2018MNRAS.477.1206N} {477, 1206}

\bibitem[\protect\citeauthoryear{{Nelson}, {Vogelsberger}, {Genel}, {Sijacki},
  {Kere{\v s}}, {Springel}  \& {Hernquist}}{{Nelson} et~al.}{2013}]{nelson13}
{Nelson} D.,  {Vogelsberger} M.,  {Genel} S.,  {Sijacki} D.,  {Kere{\v s}} D.,
  {Springel} V.,   {Hernquist} L.,  2013, \mn@doi [\mnras]
  {10.1093/mnras/sts595}, \href
  {http://adsabs.harvard.edu/abs/2013MNRAS.429.3353N} {429, 3353}

\bibitem[\protect\citeauthoryear{{Nelson} et~al.,}{{Nelson}
  et~al.}{2018a}]{nelson18a}
{Nelson} D.,  et~al., 2018a, \mn@doi [\mnras] {10.1093/mnras/stx3040}, \href
  {http://adsabs.harvard.edu/abs/2018MNRAS.475..624N} {475, 624}

\bibitem[\protect\citeauthoryear{{Nelson} et~al.,}{{Nelson}
  et~al.}{2018b}]{nelson18b}
{Nelson} D.,  et~al., 2018b, \mn@doi [\mnras] {10.1093/mnras/sty656}, \href
  {http://adsabs.harvard.edu/abs/2018MNRAS.477..450N} {477, 450}

\bibitem[\protect\citeauthoryear{{Nelson} et~al.,}{{Nelson}
  et~al.}{2019a}]{nelson19a}
{Nelson} D.,  et~al., 2019a, \mn@doi [Computational Astrophysics and Cosmology]
  {10.1186/s40668-019-0028-x}, \href
  {https://ui.adsabs.harvard.edu/abs/2019ComAC...6....2N} {6, 2}

\bibitem[\protect\citeauthoryear{{Nelson} et~al.,}{{Nelson}
  et~al.}{2019b}]{nelson19b}
{Nelson} D.,  et~al., 2019b, \mn@doi [\mnras] {10.1093/mnras/stz2306}, \href
  {https://ui.adsabs.harvard.edu/abs/2019MNRAS.490.3234N} {490, 3234}

\bibitem[\protect\citeauthoryear{{Nelson} et~al.,}{{Nelson}
  et~al.}{2020}]{nelson20}
{Nelson} D.,  et~al., 2020, arXiv e-prints, \href
  {https://ui.adsabs.harvard.edu/abs/2020arXiv200509654N} {p. arXiv:2005.09654}

\bibitem[\protect\citeauthoryear{{Nielsen}, {Churchill}, {Kacprzak}, {Murphy}
  \& {Evans}}{{Nielsen} et~al.}{2015}]{nielsen15}
{Nielsen} N.~M.,  {Churchill} C.~W.,  {Kacprzak} G.~G.,  {Murphy} M.~T.,
  {Evans} J.~L.,  2015, \mn@doi [\apj] {10.1088/0004-637X/812/1/83}, \href
  {http://adsabs.harvard.edu/abs/2015ApJ...812...83N} {812, 83}

\bibitem[\protect\citeauthoryear{{Nielsen}, {Kacprzak}, {Pointon}, {Murphy},
  {Churchill}  \& {Dav{\'e}}}{{Nielsen} et~al.}{2020}]{nielsen2020}
{Nielsen} N.~M.,  {Kacprzak} G.~G.,  {Pointon} S.~K.,  {Murphy} M.~T.,
  {Churchill} C.~W.,   {Dav{\'e}} R.,  2020, arXiv e-prints, \href
  {https://ui.adsabs.harvard.edu/abs/2020arXiv200208516N} {p. arXiv:2002.08516}

\bibitem[\protect\citeauthoryear{{Oppenheimer} \& {Dav{\'e}}}{{Oppenheimer} \&
  {Dav{\'e}}}{2006}]{oppenheimer06}
{Oppenheimer} B.~D.,  {Dav{\'e}} R.,  2006, \mn@doi [\mnras]
  {10.1111/j.1365-2966.2006.10989.x}, \href
  {https://ui.adsabs.harvard.edu/abs/2006MNRAS.373.1265O} {373, 1265}

\bibitem[\protect\citeauthoryear{{Oppenheimer} \& {Dav{\'e}}}{{Oppenheimer} \&
  {Dav{\'e}}}{2008}]{oppenheimer2008}
{Oppenheimer} B.~D.,  {Dav{\'e}} R.,  2008, \mn@doi [\mnras]
  {10.1111/j.1365-2966.2008.13280.x}, \href
  {https://ui.adsabs.harvard.edu/abs/2008MNRAS.387..577O} {387, 577}

\bibitem[\protect\citeauthoryear{{Oppenheimer} \& {Dav{\'e}}}{{Oppenheimer} \&
  {Dav{\'e}}}{2009}]{oppenheimer2009}
{Oppenheimer} B.~D.,  {Dav{\'e}} R.,  2009, \mn@doi [\mnras]
  {10.1111/j.1365-2966.2009.14676.x}, \href
  {https://ui.adsabs.harvard.edu/abs/2009MNRAS.395.1875O} {395, 1875}

\bibitem[\protect\citeauthoryear{Oppenheimer, Dav\'e, Keres, Katz, Kollmeier
  \& Weinberg}{Oppenheimer et~al.}{2010b}]{oppenheimer2010}
Oppenheimer B.,  Dav\'e R.,  Keres D.,  Katz N.,  Kollmeier J.,   Weinberg D.,
  2010b, MNRAS, 406, 2325

\bibitem[\protect\citeauthoryear{{Oppenheimer}, {Dav{\'e}}, {Kere{\v s}},
  {Fardal}, {Katz}, {Kollmeier}  \& {Weinberg}}{{Oppenheimer}
  et~al.}{2010a}]{oppenheimer10}
{Oppenheimer} B.~D.,  {Dav{\'e}} R.,  {Kere{\v s}} D.,  {Fardal} M.,  {Katz}
  N.,  {Kollmeier} J.~A.,   {Weinberg} D.~H.,  2010a, \mn@doi [\mnras]
  {10.1111/j.1365-2966.2010.16872.x}, \href
  {http://adsabs.harvard.edu/abs/2010MNRAS.406.2325O} {406, 2325}

\bibitem[\protect\citeauthoryear{{Pakmor} \& {Springel}}{{Pakmor} \&
  {Springel}}{2013}]{pakmor13}
{Pakmor} R.,  {Springel} V.,  2013, \mn@doi [\mnras] {10.1093/mnras/stt428},
  \href {http://adsabs.harvard.edu/abs/2013MNRAS.432..176P} {432, 176}

\bibitem[\protect\citeauthoryear{{Pakmor}, {Bauer}  \& {Springel}}{{Pakmor}
  et~al.}{2011}]{pakmor11}
{Pakmor} R.,  {Bauer} A.,   {Springel} V.,  2011, \mn@doi [\mnras]
  {10.1111/j.1365-2966.2011.19591.x}, \href
  {http://adsabs.harvard.edu/abs/2011MNRAS.418.1392P} {418, 1392}

\bibitem[\protect\citeauthoryear{{Peeples} et~al.,}{{Peeples}
  et~al.}{2019}]{peeples19}
{Peeples} M.~S.,  et~al., 2019, \mn@doi [\apj] {10.3847/1538-4357/ab0654},
  \href {https://ui.adsabs.harvard.edu/abs/2019ApJ...873..129P} {873, 129}

\bibitem[\protect\citeauthoryear{{P{\'e}roux} \& {Howk}}{{P{\'e}roux} \&
  {Howk}}{2020}]{peroux2020}
{P{\'e}roux} C.,  {Howk} J.~C.,  2020, \mn@doi [\araa]
  {10.1146/annurev-astro-021820-120014}, \href
  {https://ui.adsabs.harvard.edu/abs/2020ARA&A..5821820P} {58, 363}

\bibitem[\protect\citeauthoryear{P\'eroux, Bouch\'e, Kulkarni, York  \&
  Vladilo}{P\'eroux et~al.}{2011}]{peroux2011}
P\'eroux C.,  Bouch\'e N.,  Kulkarni V.,  York D.,   Vladilo G.,  2011, MNRAS,
  410, 2237

\bibitem[\protect\citeauthoryear{P\'eroux, Bouch\'e, Kulkarni, York  \&
  Vladilo}{P\'eroux et~al.}{2012}]{peroux2012}
P\'eroux C.,  Bouch\'e N.,  Kulkarni V.,  York D.,   Vladilo G.,  2012, MNRAS,
  419, 3060

\bibitem[\protect\citeauthoryear{{P{\'e}roux}, {Bouch{\'e}}, {Kulkarni}  \&
  {York}}{{P{\'e}roux} et~al.}{2013}]{peroux2013}
{P{\'e}roux} C.,  {Bouch{\'e}} N.,  {Kulkarni} V.~P.,   {York} D.~G.,  2013,
  \mn@doi [\mnras] {10.1093/mnras/stt1760}, \href
  {https://ui.adsabs.harvard.edu/abs/2013MNRAS.436.2650P} {436, 2650}

\bibitem[\protect\citeauthoryear{{P{\'e}roux} et~al.,}{{P{\'e}roux}
  et~al.}{2016}]{peroux2016}
{P{\'e}roux} C.,  et~al., 2016, \mn@doi [MNRAS] {10.1093/mnras/stw016}, \href
  {http://adsabs.harvard.edu/abs/2016MNRAS.457..903P} {457, 903}

\bibitem[\protect\citeauthoryear{{P{\'e}roux} et~al.,}{{P{\'e}roux}
  et~al.}{2017}]{peroux2017}
{P{\'e}roux} C.,  et~al., 2017, \mn@doi [MNRAS] {10.1093/mnras/stw2444}, \href
  {http://cdsads.u-strasbg.fr/abs/2017MNRAS.464.2053P} {464, 2053}

\bibitem[\protect\citeauthoryear{{P{\'e}roux} et~al.,}{{P{\'e}roux}
  et~al.}{2019}]{peroux2019}
{P{\'e}roux} C.,  et~al., 2019, \mn@doi [\mnras] {10.1093/mnras/stz202}, \href
  {https://ui.adsabs.harvard.edu/\#abs/2019MNRAS.485.1595P} {485, 1595}

\bibitem[\protect\citeauthoryear{Pettini, Shapley, Steidel, Cuby, Dickinson,
  Moorwood, Adelberger  \& Giavalisco}{Pettini et~al.}{2001}]{pettini2001}
Pettini M.,  Shapley A.,  Steidel C.,  Cuby J.,  Dickinson M.,  Moorwood A.,
  Adelberger K.,   Giavalisco M.,  2001, ApJ, 554, 981

\bibitem[\protect\citeauthoryear{{Pillepich} et~al.,}{{Pillepich}
  et~al.}{2018a}]{pillepich18a}
{Pillepich} A.,  et~al., 2018a, \mn@doi [\mnras] {10.1093/mnras/stx2656}, \href
  {http://adsabs.harvard.edu/abs/2018MNRAS.473.4077P} {473, 4077}

\bibitem[\protect\citeauthoryear{{Pillepich} et~al.,}{{Pillepich}
  et~al.}{2018b}]{pillepich18b}
{Pillepich} A.,  et~al., 2018b, \mn@doi [\mnras] {10.1093/mnras/stx3112}, \href
  {http://adsabs.harvard.edu/abs/2018MNRAS.475..648P} {475, 648}

\bibitem[\protect\citeauthoryear{{Pillepich} et~al.,}{{Pillepich}
  et~al.}{2019}]{pillepich19}
{Pillepich} A.,  et~al., 2019, \mn@doi [\mnras] {10.1093/mnras/stz2338}, \href
  {https://ui.adsabs.harvard.edu/abs/2019MNRAS.490.3196P} {490, 3196}

\bibitem[\protect\citeauthoryear{{Ploeckinger} \& {Schaye}}{{Ploeckinger} \&
  {Schaye}}{2020}]{ploeckinger20}
{Ploeckinger} S.,  {Schaye} J.,  2020, arXiv e-prints, \href
  {https://ui.adsabs.harvard.edu/abs/2020arXiv200614322P} {p. arXiv:2006.14322}

\bibitem[\protect\citeauthoryear{{Pointon}, {Kacprzak}, {Nielsen}, {Muzahid},
  {Murphy}, {Churchill}  \& {Charlton}}{{Pointon} et~al.}{2019}]{pointon2019}
{Pointon} S.~K.,  {Kacprzak} G.~G.,  {Nielsen} N.~M.,  {Muzahid} S.,  {Murphy}
  M.~T.,  {Churchill} C.~W.,   {Charlton} J.~C.,  2019, \mn@doi [\apj]
  {10.3847/1538-4357/ab3b0e}, \href
  {https://ui.adsabs.harvard.edu/abs/2019ApJ...883...78P} {883, 78}

\bibitem[\protect\citeauthoryear{{Poudel}, {Kulkarni}, {Morrison},
  {P{\'e}roux}, {Som}, {Rahmani}  \& {Quiret}}{{Poudel}
  et~al.}{2018}]{poudel2018}
{Poudel} S.,  {Kulkarni} V.~P.,  {Morrison} S.,  {P{\'e}roux} C.,  {Som} D.,
  {Rahmani} H.,   {Quiret} S.,  2018, \mn@doi [\mnras] {10.1093/mnras/stx2607},
  \href {https://ui.adsabs.harvard.edu/#abs/2018MNRAS.473.3559P} {473, 3559}

\bibitem[\protect\citeauthoryear{{Prochaska}, {Hennawi}  \&
  {Simcoe}}{{Prochaska} et~al.}{2013}]{prochaska13}
{Prochaska} J.~X.,  {Hennawi} J.~F.,   {Simcoe} R.~A.,  2013, \mn@doi [\apjl]
  {10.1088/2041-8205/762/2/L19}, \href
  {http://adsabs.harvard.edu/abs/2013ApJ...762L..19P} {762, L19}

\bibitem[\protect\citeauthoryear{{Prochaska} et~al.,}{{Prochaska}
  et~al.}{2017}]{prochaska2017}
{Prochaska} J.~X.,  et~al., 2017, \mn@doi [\apj] {10.3847/1538-4357/aa6007},
  \href {https://ui.adsabs.harvard.edu/abs/2017ApJ...837..169P} {837, 169}

\bibitem[\protect\citeauthoryear{{Quiret} et~al.,}{{Quiret}
  et~al.}{2016}]{quiret2016}
{Quiret} S.,  et~al., 2016, \mn@doi [\mnras] {10.1093/mnras/stw524}, \href
  {http://adsabs.harvard.edu/abs/2016MNRAS.458.4074Q} {458, 4074}

\bibitem[\protect\citeauthoryear{{Rahmani} et~al.,}{{Rahmani}
  et~al.}{2018a}]{rahmani2018a}
{Rahmani} H.,  et~al., 2018a, \mn@doi [\mnras] {10.1093/mnras/stx2726}, \href
  {https://ui.adsabs.harvard.edu/abs/2018MNRAS.474..254R} {474, 254}

\bibitem[\protect\citeauthoryear{{Rahmani} et~al.,}{{Rahmani}
  et~al.}{2018b}]{rahmani2018b}
{Rahmani} H.,  et~al., 2018b, \mn@doi [\mnras] {10.1093/mnras/sty2216}, \href
  {https://ui.adsabs.harvard.edu/abs/2018MNRAS.480.5046R} {480, 5046}

\bibitem[\protect\citeauthoryear{{Rahmati} \& {Oppenheimer}}{{Rahmati} \&
  {Oppenheimer}}{2018}]{rahmati2018}
{Rahmati} A.,  {Oppenheimer} B.~D.,  2018, \mn@doi [\mnras]
  {10.1093/mnras/sty610}, \href
  {https://ui.adsabs.harvard.edu/abs/2018MNRAS.476.4865R} {476, 4865}

\bibitem[\protect\citeauthoryear{{Rahmati}, {Schaye}, {Bower}, {Crain},
  {Furlong}, {Schaller}  \& {Theuns}}{{Rahmati} et~al.}{2015}]{rahmati2015}
{Rahmati} A.,  {Schaye} J.,  {Bower} R.~G.,  {Crain} R.~A.,  {Furlong} M.,
  {Schaller} M.,   {Theuns} T.,  2015, \mn@doi [\mnras]
  {10.1093/mnras/stv1414}, \href
  {https://ui.adsabs.harvard.edu/abs/2015MNRAS.452.2034R} {452, 2034}

\bibitem[\protect\citeauthoryear{{Rahmati}, {Schaye}, {Crain}, {Oppenheimer},
  {Schaller}  \& {Theuns}}{{Rahmati} et~al.}{2016}]{rahmati16}
{Rahmati} A.,  {Schaye} J.,  {Crain} R.~A.,  {Oppenheimer} B.~D.,  {Schaller}
  M.,   {Theuns} T.,  2016, \mn@doi [\mnras] {10.1093/mnras/stw453}, \href
  {http://adsabs.harvard.edu/abs/2016MNRAS.459..310R} {459, 310}

\bibitem[\protect\citeauthoryear{{Ribaudo}, {Lehner}, {Howk}, {Werk}, {Tripp},
  {Prochaska}, {Meiring}  \& {Tumlinson}}{{Ribaudo} et~al.}{2011}]{ribaudo2011}
{Ribaudo} J.,  {Lehner} N.,  {Howk} J.~C.,  {Werk} J.~K.,  {Tripp} T.~M.,
  {Prochaska} J.~X.,  {Meiring} J.~D.,   {Tumlinson} J.,  2011, \mn@doi [\apj]
  {10.1088/0004-637X/743/2/207}, \href
  {https://ui.adsabs.harvard.edu/abs/2011ApJ...743..207R} {743, 207}

\bibitem[\protect\citeauthoryear{{Rosas-Guevara} et~al.,}{{Rosas-Guevara}
  et~al.}{2015}]{rosas2015}
{Rosas-Guevara} Y.~M.,  et~al., 2015, \mn@doi [\mnras] {10.1093/mnras/stv2056},
  \href {https://ui.adsabs.harvard.edu/abs/2015MNRAS.454.1038R} {454, 1038}

\bibitem[\protect\citeauthoryear{{Rubin}, {Prochaska}, {Koo}  \&
  {Phillips}}{{Rubin} et~al.}{2012}]{rubin2012}
{Rubin} K. H.~R.,  {Prochaska} J.~X.,  {Koo} D.~C.,   {Phillips} A.~C.,  2012,
  \mn@doi [\apjl] {10.1088/2041-8205/747/2/L26}, \href
  {https://ui.adsabs.harvard.edu/abs/2012ApJ...747L..26R} {747, L26}

\bibitem[\protect\citeauthoryear{{Rubin}, {Prochaska}, {Koo}, {Phillips},
  {Martin}  \& {Winstrom}}{{Rubin} et~al.}{2014}]{rubin2014}
{Rubin} K. H.~R.,  {Prochaska} J.~X.,  {Koo} D.~C.,  {Phillips} A.~C.,
  {Martin} C.~L.,   {Winstrom} L.~O.,  2014, \mn@doi [\apj]
  {10.1088/0004-637X/794/2/156}, \href
  {https://ui.adsabs.harvard.edu/abs/2014ApJ...794..156R} {794, 156}

\bibitem[\protect\citeauthoryear{{Rubin}, {Diamond-Stanic}, {Coil}, {Crighton}
  \& {Moustakas}}{{Rubin} et~al.}{2018}]{rubin18}
{Rubin} K.~H.~R.,  {Diamond-Stanic} A.~M.,  {Coil} A.~L.,  {Crighton} N.~H.~M.,
    {Moustakas} J.,  2018, \mn@doi [\apj] {10.3847/1538-4357/aa9792}, \href
  {http://adsabs.harvard.edu/abs/2018ApJ...853...95R} {853, 95}

\bibitem[\protect\citeauthoryear{{Rudie}, {Steidel}, {Pettini}, {Trainor},
  {Strom}, {Hummels}, {Reddy}  \& {Shapley}}{{Rudie} et~al.}{2019}]{rudie2019}
{Rudie} G.~C.,  {Steidel} C.~C.,  {Pettini} M.,  {Trainor} R.~F.,  {Strom}
  A.~L.,  {Hummels} C.~B.,  {Reddy} N.~A.,   {Shapley} A.~E.,  2019, \mn@doi
  [\apj] {10.3847/1538-4357/ab4255}, \href
  {https://ui.adsabs.harvard.edu/abs/2019ApJ...885...61R} {885, 61}

\bibitem[\protect\citeauthoryear{{Scannapieco} \& {Br{\"u}ggen}}{{Scannapieco}
  \& {Br{\"u}ggen}}{2015}]{scannapieco15}
{Scannapieco} E.,  {Br{\"u}ggen} M.,  2015, \mn@doi [\apj]
  {10.1088/0004-637X/805/2/158}, \href
  {http://adsabs.harvard.edu/abs/2015ApJ...805..158S} {805, 158}

\bibitem[\protect\citeauthoryear{{Schaye} \& {Dalla Vecchia}}{{Schaye} \&
  {Dalla Vecchia}}{2008}]{Schaye2008}
{Schaye} J.,  {Dalla Vecchia} C.,  2008, \mn@doi [\mnras]
  {10.1111/j.1365-2966.2007.12639.x}, \href
  {https://ui.adsabs.harvard.edu/abs/2008MNRAS.383.1210S} {383, 1210}

\bibitem[\protect\citeauthoryear{{Schaye} et~al.,}{{Schaye}
  et~al.}{2015}]{schaye2015}
{Schaye} J.,  et~al., 2015, \mn@doi [\mnras] {10.1093/mnras/stu2058}, \href
  {https://ui.adsabs.harvard.edu/\#abs/2015MNRAS.446..521S} {446, 521}

\bibitem[\protect\citeauthoryear{Schroetter, Bouch\'e, P\'eroux, Murphy,
  Contini  \& Finley}{Schroetter et~al.}{2015}]{schroetter2015}
Schroetter I.,  Bouch\'e N.,  P\'eroux C.,  Murphy M.,  Contini T.,   Finley
  H.,  2015, ApJ, 804, 83

\bibitem[\protect\citeauthoryear{{Schroetter} et~al.,}{{Schroetter}
  et~al.}{2016}]{schroetter2016}
{Schroetter} I.,  et~al., 2016, \mn@doi [ApJ] {10.3847/1538-4357/833/1/39},
  \href {http://adsabs.harvard.edu/abs/2016ApJ...833...39S} {833, 39}

\bibitem[\protect\citeauthoryear{{Schroetter} et~al.,}{{Schroetter}
  et~al.}{2019}]{schroetter2019}
{Schroetter} I.,  et~al., 2019, \mn@doi [\mnras] {10.1093/mnras/stz2822}, \href
  {https://ui.adsabs.harvard.edu/abs/2019MNRAS.tmp.2451S} {p.~2451}

\bibitem[\protect\citeauthoryear{Shapley, Steidel, Pettini  \&
  Adelberger}{Shapley et~al.}{2003}]{shapley2003}
Shapley a.,  Steidel C.,  Pettini M.,   Adelberger K.,  2003, ApJ, 588, 65

\bibitem[\protect\citeauthoryear{{Shen}, {Wadsley}  \& {Stinson}}{{Shen}
  et~al.}{2010}]{shen10}
{Shen} S.,  {Wadsley} J.,   {Stinson} G.,  2010, \mn@doi [\mnras]
  {10.1111/j.1365-2966.2010.17047.x}, \href
  {https://ui.adsabs.harvard.edu/abs/2010MNRAS.407.1581S} {407, 1581}

\bibitem[\protect\citeauthoryear{{Shen}, {Madau}, {Aguirre}, {Guedes}, {Mayer}
  \& {Wadsley}}{{Shen} et~al.}{2012}]{shen2012}
{Shen} S.,  {Madau} P.,  {Aguirre} A.,  {Guedes} J.,  {Mayer} L.,   {Wadsley}
  J.,  2012, \mn@doi [\apj] {10.1088/0004-637X/760/1/50}, \href
  {https://ui.adsabs.harvard.edu/abs/2012ApJ...760...50S} {760, 50}

\bibitem[\protect\citeauthoryear{{Shen}, {Madau}, {Guedes}, {Mayer},
  {Prochaska}  \& {Wadsley}}{{Shen} et~al.}{2013}]{shen2013}
{Shen} S.,  {Madau} P.,  {Guedes} J.,  {Mayer} L.,  {Prochaska} J.~X.,
  {Wadsley} J.,  2013, \mn@doi [\apj] {10.1088/0004-637X/765/2/89}, \href
  {https://ui.adsabs.harvard.edu/abs/2013ApJ...765...89S} {765, 89}

\bibitem[\protect\citeauthoryear{{Silk}}{{Silk}}{1977}]{silk77}
{Silk} J.,  1977, \mn@doi [\apj] {10.1086/154972}, \href
  {http://adsabs.harvard.edu/abs/1977ApJ...211..638S} {211, 638}

\bibitem[\protect\citeauthoryear{{Somerville}, {Popping}  \&
  {Trager}}{{Somerville} et~al.}{2015}]{somerville2015}
{Somerville} R.~S.,  {Popping} G.,   {Trager} S.~C.,  2015, \mn@doi [\mnras]
  {10.1093/mnras/stv1877}, \href
  {https://ui.adsabs.harvard.edu/abs/2015MNRAS.453.4337S} {453, 4337}

\bibitem[\protect\citeauthoryear{{Springel}}{{Springel}}{2010}]{spr10}
{Springel} V.,  2010, \mn@doi [\mnras] {10.1111/j.1365-2966.2009.15715.x}, 401,
  791

\bibitem[\protect\citeauthoryear{{Springel} \& {Hernquist}}{{Springel} \&
  {Hernquist}}{2003}]{springel03}
{Springel} V.,  {Hernquist} L.,  2003, \mn@doi [\mnras]
  {10.1046/j.1365-8711.2003.06206.x}, \href
  {https://ui.adsabs.harvard.edu/abs/2003MNRAS.339..289S} {339, 289}

\bibitem[\protect\citeauthoryear{{Springel}, {White}, {Tormen}  \&
  {Kauffmann}}{{Springel} et~al.}{2001}]{spr01}
{Springel} V.,  {White} S.~D.~M.,  {Tormen} G.,   {Kauffmann} G.,  2001,
  \mn@doi [\mnras] {10.1046/j.1365-8711.2001.04912.x}, \href
  {http://adsabs.harvard.edu/abs/2001MNRAS.328..726S} {328, 726}

\bibitem[\protect\citeauthoryear{{Springel}, {Di Matteo}  \&
  {Hernquist}}{{Springel} et~al.}{2005a}]{springel05a}
{Springel} V.,  {Di Matteo} T.,   {Hernquist} L.,  2005a, \mn@doi [\mnras]
  {10.1111/j.1365-2966.2005.09238.x}, \href
  {https://ui.adsabs.harvard.edu/abs/2005MNRAS.361..776S} {361, 776}

\bibitem[\protect\citeauthoryear{{Springel} et~al.,}{{Springel}
  et~al.}{2005b}]{springel2005}
{Springel} V.,  et~al., 2005b, \mn@doi [\nat] {10.1038/nature03597}, \href
  {http://cdsads.u-strasbg.fr/abs/2005Natur.435..629S} {435, 629}

\bibitem[\protect\citeauthoryear{{Springel} et~al.,}{{Springel}
  et~al.}{2018}]{springel18}
{Springel} V.,  et~al., 2018, \mn@doi [\mnras] {10.1093/mnras/stx3304}, \href
  {http://adsabs.harvard.edu/abs/2018MNRAS.475..676S} {475, 676}

\bibitem[\protect\citeauthoryear{Steidel, Erb, Shapley, Pettini, Reddy,
  Bogosavljevic, Rudie  \& Rakic}{Steidel et~al.}{2010}]{steidel2010}
Steidel C.,  Erb D.,  Shapley A.,  Pettini M.,  Reddy N.,  Bogosavljevic M.,
  Rudie G.,   Rakic O.,  2010, ApJ, 717, 289

\bibitem[\protect\citeauthoryear{Stewart, Kaufmann, Bullock, Barton, Maller,
  Biemand  \& Wadsley}{Stewart et~al.}{2011}]{stewart2011}
Stewart K.,  Kaufmann T.,  Bullock J.,  Barton E.,  Maller A.,  Biemand J.,
  Wadsley J.,  2011, ApJ, 738, 39

\bibitem[\protect\citeauthoryear{{Suresh}, {Nelson}, {Genel}, {Rubin}  \&
  {Hernquist}}{{Suresh} et~al.}{2019}]{suresh19}
{Suresh} J.,  {Nelson} D.,  {Genel} S.,  {Rubin} K. H.~R.,   {Hernquist} L.,
  2019, \mn@doi [\mnras] {10.1093/mnras/sty3402}, \href
  {https://ui.adsabs.harvard.edu/abs/2019MNRAS.483.4040S} {483, 4040}

\bibitem[\protect\citeauthoryear{{Torrey} et~al.,}{{Torrey}
  et~al.}{2019}]{torrey19}
{Torrey} P.,  et~al., 2019, \mn@doi [\mnras] {10.1093/mnras/stz243}, \href
  {https://ui.adsabs.harvard.edu/abs/2019MNRAS.484.5587T} {484, 5587}

\bibitem[\protect\citeauthoryear{{Tripp} et~al.,}{{Tripp}
  et~al.}{2011}]{tripp2011}
{Tripp} T.~M.,  et~al., 2011, \mn@doi [Science] {10.1126/science.1209850},
  \href {https://ui.adsabs.harvard.edu/abs/2011Sci...334..952T} {334, 952}

\bibitem[\protect\citeauthoryear{{Tumlinson}, {Peeples}  \& {Werk}}{{Tumlinson}
  et~al.}{2017}]{tumlinson2017}
{Tumlinson} J.,  {Peeples} M.~S.,   {Werk} J.~K.,  2017, \mn@doi [\araa]
  {10.1146/annurev-astro-091916-055240}, \href
  {https://ui.adsabs.harvard.edu/abs/2017ARA&A..55..389T} {55, 389}

\bibitem[\protect\citeauthoryear{{Turner}, {Schaye}, {Crain}, {Theuns}  \&
  {Wendt}}{{Turner} et~al.}{2016}]{turner16}
{Turner} M.~L.,  {Schaye} J.,  {Crain} R.~A.,  {Theuns} T.,   {Wendt} M.,
  2016, \mn@doi [\mnras] {10.1093/mnras/stw1816}, \href
  {http://adsabs.harvard.edu/abs/2016MNRAS.462.2440T} {462, 2440}

\bibitem[\protect\citeauthoryear{{Veilleux}, {Cecil}  \&
  {Bland-Hawthorn}}{{Veilleux} et~al.}{2005}]{veilleux2005}
{Veilleux} S.,  {Cecil} G.,   {Bland-Hawthorn} J.,  2005, \mn@doi [\araa]
  {10.1146/annurev.astro.43.072103.150610}, \href
  {https://ui.adsabs.harvard.edu/abs/2005ARA&A..43..769V} {43, 769}

\bibitem[\protect\citeauthoryear{{Vogelsberger} et~al.,}{{Vogelsberger}
  et~al.}{2014}]{vog14a}
{Vogelsberger} M.,  et~al., 2014, \mn@doi [\nat] {10.1038/nature13316}, \href
  {http://adsabs.harvard.edu/abs/2014Natur.509..177V} {509, 177}

\bibitem[\protect\citeauthoryear{{Weinberger} et~al.,}{{Weinberger}
  et~al.}{2017}]{weinberger17}
{Weinberger} R.,  et~al., 2017, \mn@doi [\mnras] {10.1093/mnras/stw2944}, \href
  {http://adsabs.harvard.edu/abs/2017MNRAS.465.3291W} {465, 3291}

\bibitem[\protect\citeauthoryear{{Weinberger} et~al.,}{{Weinberger}
  et~al.}{2018}]{weinberger18}
{Weinberger} R.,  et~al., 2018, \mn@doi [\mnras] {10.1093/mnras/sty1733}, \href
  {http://adsabs.harvard.edu/abs/2018MNRAS.479.4056W} {479, 4056}

\bibitem[\protect\citeauthoryear{Weiner et~al.,}{Weiner
  et~al.}{2009}]{weiner2009}
Weiner B.,  et~al., 2009, ApJ, 692, 187

\bibitem[\protect\citeauthoryear{{Werk} et~al.,}{{Werk}
  et~al.}{2014}]{werk2014}
{Werk} J.~K.,  et~al., 2014, \mn@doi [ApJ] {10.1088/0004-637X/792/1/8}, \href
  {http://adsabs.harvard.edu/abs/2014ApJ...792....8W} {792, 8}

\bibitem[\protect\citeauthoryear{{White} \& {Rees}}{{White} \&
  {Rees}}{1978}]{wr78}
{White} S.~D.~M.,  {Rees} M.~J.,  1978, \mnras, \href
  {http://adsabs.harvard.edu/abs/1978MNRAS.183..341W} {183, 341}

\bibitem[\protect\citeauthoryear{{Wiersma}, {Schaye}  \& {Smith}}{{Wiersma}
  et~al.}{2009a}]{wiersma09a}
{Wiersma} R.~P.~C.,  {Schaye} J.,   {Smith} B.~D.,  2009a, \mn@doi [\mnras]
  {10.1111/j.1365-2966.2008.14191.x}, \href
  {http://adsabs.harvard.edu/abs/2009MNRAS.393...99W} {393, 99}

\bibitem[\protect\citeauthoryear{{Wiersma}, {Schaye}, {Theuns}, {Dalla Vecchia}
   \& {Tornatore}}{{Wiersma} et~al.}{2009b}]{wiersma2009}
{Wiersma} R. P.~C.,  {Schaye} J.,  {Theuns} T.,  {Dalla Vecchia} C.,
  {Tornatore} L.,  2009b, \mn@doi [\mnras] {10.1111/j.1365-2966.2009.15331.x},
  \href {https://ui.adsabs.harvard.edu/abs/2009MNRAS.399..574W} {399, 574}

\bibitem[\protect\citeauthoryear{{Wiersma}, {Schaye}, {Theuns}, {Dalla Vecchia}
   \& {Tornatore}}{{Wiersma} et~al.}{2009c}]{wiersma09b}
{Wiersma} R. P.~C.,  {Schaye} J.,  {Theuns} T.,  {Dalla Vecchia} C.,
  {Tornatore} L.,  2009c, \mn@doi [\mnras] {10.1111/j.1365-2966.2009.15331.x},
  \href {https://ui.adsabs.harvard.edu/abs/2009MNRAS.399..574W} {399, 574}

\bibitem[\protect\citeauthoryear{{Wijers}, {Schaye}, {Oppenheimer}, {Crain}  \&
  {Nicastro}}{{Wijers} et~al.}{2019}]{wijers2019}
{Wijers} N.~A.,  {Schaye} J.,  {Oppenheimer} B.~D.,  {Crain} R.~A.,
  {Nicastro} F.,  2019, \mn@doi [\mnras] {10.1093/mnras/stz1762}, \href
  {https://ui.adsabs.harvard.edu/abs/2019MNRAS.488.2947W} {488, 2947}

\bibitem[\protect\citeauthoryear{{Wotta}, {Lehner}, {Howk}, {O'Meara}  \&
  {Prochaska}}{{Wotta} et~al.}{2016}]{wotta2016}
{Wotta} C.~B.,  {Lehner} N.,  {Howk} J.~C.,  {O'Meara} J.~M.,   {Prochaska}
  J.~X.,  2016, \mn@doi [\apj] {10.3847/0004-637X/831/1/95}, \href
  {http://adsabs.harvard.edu/abs/2016ApJ...831...95W} {831, 95}

\bibitem[\protect\citeauthoryear{{Wotta}, {Lehner}, {Howk}, {O'Meara},
  {Oppenheimer}  \& {Cooksey}}{{Wotta} et~al.}{2019}]{wotta2019}
{Wotta} C.~B.,  {Lehner} N.,  {Howk} J.~C.,  {O'Meara} J.~M.,  {Oppenheimer}
  B.~D.,   {Cooksey} K.~L.,  2019, \mn@doi [\apj] {10.3847/1538-4357/aafb74},
  \href {https://ui.adsabs.harvard.edu/abs/2019ApJ...872...81W} {872, 81}

\bibitem[\protect\citeauthoryear{{Zabl} et~al.,}{{Zabl}
  et~al.}{2019}]{zabl2019}
{Zabl} J.,  et~al., 2019, \mn@doi [\mnras] {10.1093/mnras/stz392}, \href
  {https://ui.adsabs.harvard.edu/abs/2019MNRAS.485.1961Z} {485, 1961}

\bibitem[\protect\citeauthoryear{{Zahedy}, {Chen}, {Johnson}, {Pierce},
  {Rauch}, {Huang}, {Weiner}  \& {Gauthier}}{{Zahedy}
  et~al.}{2019}]{zahedy2019}
{Zahedy} F.~S.,  {Chen} H.-W.,  {Johnson} S.~D.,  {Pierce} R.~M.,  {Rauch} M.,
  {Huang} Y.-H.,  {Weiner} B.~J.,   {Gauthier} J.-R.,  2019, \mn@doi [\mnras]
  {10.1093/mnras/sty3482}, \href
  {https://ui.adsabs.harvard.edu/abs/2019MNRAS.484.2257Z} {484, 2257}

\bibitem[\protect\citeauthoryear{{Zheng}, {Peek}, {Werk}  \& {Putman}}{{Zheng}
  et~al.}{2017}]{zheng2017}
{Zheng} Y.,  {Peek} J.~E.~G.,  {Werk} J.~K.,   {Putman} M.~E.,  2017, \mn@doi
  [\apj] {10.3847/1538-4357/834/2/179}, \href
  {https://ui.adsabs.harvard.edu/abs/2017ApJ...834..179Z} {834, 179}

\bibitem[\protect\citeauthoryear{{Zhu} et~al.,}{{Zhu} et~al.}{2014}]{zhu14}
{Zhu} G.,  et~al., 2014, \mn@doi [\mnras] {10.1093/mnras/stu186}, \href
  {https://ui.adsabs.harvard.edu/abs/2014MNRAS.439.3139Z} {439, 3139}

\bibitem[\protect\citeauthoryear{{van de Voort} \& {Schaye}}{{van de Voort} \&
  {Schaye}}{2012}]{vandevoort2012b}
{van de Voort} F.,  {Schaye} J.,  2012, \mn@doi [\mnras]
  {10.1111/j.1365-2966.2012.20949.x}, \href
  {https://ui.adsabs.harvard.edu/abs/2012MNRAS.423.2991V} {423, 2991}

\bibitem[\protect\citeauthoryear{{van de Voort} \& {Schaye}}{{van de Voort} \&
  {Schaye}}{2013}]{vandevoort2013}
{van de Voort} F.,  {Schaye} J.,  2013, \mn@doi [\mnras]
  {10.1093/mnras/stt115}, \href
  {https://ui.adsabs.harvard.edu/abs/2013MNRAS.430.2688V} {430, 2688}

\bibitem[\protect\citeauthoryear{{van de Voort}, {Schaye}, {Booth}, {Haas}  \&
  {Dalla Vecchia}}{{van de Voort} et~al.}{2011}]{vandevoort2011a}
{van de Voort} F.,  {Schaye} J.,  {Booth} C.~M.,  {Haas} M.~R.,   {Dalla
  Vecchia} C.,  2011, \mn@doi [\mnras] {10.1111/j.1365-2966.2011.18565.x},
  \href {https://ui.adsabs.harvard.edu/abs/2011MNRAS.414.2458V} {414, 2458}

\bibitem[\protect\citeauthoryear{{van de Voort}, {Schaye}, {Altay}  \&
  {Theuns}}{{van de Voort} et~al.}{2012}]{vandevoort2012a}
{van de Voort} F.,  {Schaye} J.,  {Altay} G.,   {Theuns} T.,  2012, \mn@doi
  [\mnras] {10.1111/j.1365-2966.2012.20487.x}, \href
  {https://ui.adsabs.harvard.edu/abs/2012MNRAS.421.2809V} {421, 2809}

\bibitem[\protect\citeauthoryear{{van de Voort}, {Springel}, {Mandelker}, {van
  den Bosch}  \& {Pakmor}}{{van de Voort} et~al.}{2019}]{vandevoort2019}
{van de Voort} F.,  {Springel} V.,  {Mandelker} N.,  {van den Bosch} F.~C.,
  {Pakmor} R.,  2019, \mn@doi [\mnras] {10.1093/mnrasl/sly190}, \href
  {https://ui.adsabs.harvard.edu/abs/2019MNRAS.482L..85V} {482, L85}

\makeatother
\end{thebibliography}

\end{document}